\documentclass[pra,twocolumn]{revtex4}
\usepackage{graphicx}
\usepackage{dcolumn}
\usepackage{amsmath}

\def\be#1{\begin{equation}\label{#1}}
\def\ee{\end{equation}}
\def\bea#1{\begin{eqnarray}\label{#1}}
\def\eea{\end{eqnarray}}
\def\Bea{\begin{eqnarray*}}
\def\Eea{\end{eqnarray*}}
\def\sp{\hspace{.5em}}
\def\sph{\hspace{.25em}}
\def\spp{\sp\sp}
\def\Eq#1{Eq.(\ref{#1})}
\def\Fig#1{Fig.(\ref{#1})}

\def\no{\nonumber \\}
\def\tbf#1{\textbf{#1}}
\def\trm#1{\textrm{#1}}
\def\tit#1{\textit{#1}}

\def\mbf#1{\mbox{{\boldmath $#1$}}}
\def\half{\frac{1}{2}}
\def\bra#1{\langle#1|}
\def\ket#1{|#1\rangle}

\def\expm2piOmega{e^{-2\pi\Omega}}

\def\sp{\hspace{.25em}}
\def\d#1#2{d^{\sp(#1)}_{\Omega,#2\vec{k}_\perp}}

\def\b#1#2{b^{\sp(#1)}_{\Omega,#2\vec{k}_\perp}}

\def\a#1{a_{#1\vec{k}_\perp,k^3}}

\def\bfe#1{\mathbf{e}_{\hat{#1}}}

\def\e#1#2{e_{\hat{#1}}^{\spp #2}}
\def\inve#1#2{e_{#1}^{\spp \hat{#2}}}
\def\inveT#1#2{e^{\hat{#1}}_{\spp #2}}

\def\ha{\hat{a}}
\def\hb{\hat{b}}
\def\hc{\hat{c}}
\def\hd{\hat{d}}
\def\hi{\hat{i}}

\def\h0{\hat{0}}
\def\ho{\hat{0}}
\def\h1{\hat{1}}
\def\h2{\hat{2}}
\def\h3{\hat{3}}
\def\h4{\hat{4}}

\def\a{\alpha}
\def\b{\beta}
\def\c{\gamma}
\def\d{\delta}

\def\l{\lambda}
\def\s{\sigma}
\def\u{\mu}
\def\v{\nu}

\def\pd#1{\partial_{#1}}
\def\O{{\mathcal{O}}}

\def\m#1{\mathbf{#1}}
\def\n#1{\nabla_{\mathbf{#1}}}

\begin{document}
\bibliographystyle{apsrev}

\preprint{}

\title{The Wigner rotation for photons in an arbitrary gravitational field}



\author{P.M. Alsing$^{\dagger}$}
\email{paul.alsing@kirtland.af.mil}
\thanks{Corresponding author}%
\affiliation{$^*$Air Force Research Laboratory, Space Vehicles Directorate \\
3550 Aberdeen Ave, SE, Kirtland AFB, New Mexico, 87117-5776}
\author{G.J. Stephenson Jr}
\email{gjs@swcp.com}
\affiliation{$^\dagger$Department of Physics and Astronomy, University of New
Mexico, Albuquerque, NM 87131}

\date{\today\; Version: v3}

\begin{abstract}
We investigate the Wigner rotation for photons, which governs the change in the polarization of the photon
as it propagates through an arbitrary gravitational field.
We give explicit examples in Schwarzschild spacetime, and compare with the corresponding flat spacetime
results, which by the equivalence principle, holds locally at each spacetime point.
We discuss the implications of the Wigner rotation for entangled photon states in curved spacetime, and
lastly develop a sufficient condition for special (Fermi-Walker) frames in which the observer would detect no
Wigner rotation.
\end{abstract}

\maketitle





%
%


\section{Introduction}
\label{intro}
Recently, there has been much interest in the study of entanglement
for moving observers,  both for constant velocity observers (special relativity - SR)
and for arbitrarily accelerated observers (general relativity - GR). An excellent, recent review can
be found in Peres and Terno \cite{peres_terno} (and references therein). In SR and GR the important
ingredient that determines the description of moving states by moving observers is how such states
transform under the symmetries that govern the underlying flat or curved spacetime.
The relevant concept is that of the \tit{Wigner rotation} \cite{weinberg_qft1}, which for massive particles,
mixes up the spin components (along a given quantization axis) for a particle of definite spin by an $O(3)$ rotation,
and for massless particles, introduces a phase factor which is the product of a Wigner rotation angle
times the helicity of the state. In this paper we investigate the transformation of photon states as they traverse trajectories in an arbitrary curved spacetime (CST), and investigate the implications for the evolution of entangled states. In a companion article \cite{alsing_wigrot_spin}, similar investigations were carried out
on the role of the Wigner rotation on the entanglement of massive spin $\half$ particles in CST.

This paper is organized as follows. In Section II we review the transformation of quantum mechanical
states of massive particles under Lorentz transformations, in both flat and curved spacetime, and the
consequences for entangled states. In Section III we review the current research into similar
investigations for photons in flat spacetime. In Section IV we generalize the flat spacetime
results of the previous section to CST and give specific examples of the Wigner rotation
in the spherically symmetric Schwarzschild spacetime. In Section V we derive sufficient conditions
for the existence of reference frames for observers to measure a null Wigner rotation. In Section VI
we consider the consequences of the Wigner rotation on the entanglement of photon states and
photon wavepackets in CST. In Section VII we present a summary and our conclusions. In the appendix
we review the effect of the Wigner rotation on the rotation of the photon polarization in the
plane perpendicular to its propagation direction in flat spacetime. By Einstien's equivalence principle (EP),
which states that SR applies in the locally flat (Lorentz) tangent plane to a point $x$ in CST, the flat spacetime
examples presented have relevance when the Wigner rotation is generalized to photons and arbitrarily moving
observers in CST.

\section{Massive Particles with Spin}
\subsection{Flat Spacetime}
In quantum field theory,
the merger of quantum mechanics with SR, the particle states for massive particles
are defined by their spin (as in non-relativistic
mechanics) in the particle's rest frame, and additionally by their momentum. These two quantities
are the Casimir invariants of the ten parameter Poincare group of SR which describes ordinary rotations, boosts
and translations. The positive energy, single particle states thus form the state space for a
representation of the Poincare group. For massive particles with spin $j$,
the states are given by $\ket{\vec{p},\sigma}$, where $\vec{p}$ is the spatial portion
of the particle's 4-momentum $p^\mu$, and $\sigma\in\{-j,-j+1,\ldots,j\}$ are the components of the particle's spin
along a quantization axis in the rest frame of the particle. For massless particles, the states are given
by $\ket{\vec{p},\lambda}$ where $\lambda$ indicates the helicity states of the particle ($\lambda = \pm 1$ for
photons, $\lambda = \pm 1/2$ for massless fermions).

Under a Lorentz transformation (LT) $\mbf{\Lambda}$ the single-particle state for a massive particle
transforms under the unitrary transformation
$U(\Lambda)$ as \cite{weinberg_qft1}
\be{1}
U(\Lambda) \ket{\vec{p},\sigma} = \sum_{\sigma'} \,
D^{(j)}_{\sigma'\sigma}(W(\Lambda,\vec{p}))\,\ket{\overrightarrow{\Lambda p},\sigma'},
\ee
where $\overrightarrow{\Lambda p}$ are the spatial components of the Lorentz transformed 4-momentum, i.e.
$\vec{p}^{\sph '}$ where $p^{'\u} = \Lambda^\u_{\spp \v} \, p^\v$.
In \Eq{1}, $D^{j}_{\sigma'\sigma}(W(\Lambda,\vec{p}))$ is a $(2 j+1)\times (2 j+1)$ matrix
spinor representation of the rotation group $O(3)$, and $W(\Lambda,\vec{p})$ is called the Wigner rotation angle.
The explicit form of the Wigner rotation in matrix form is given by
\be{2}
 \mbf{W}(\Lambda,\vec{p}) =  \mbf{L}^{-1}(\Lambda p) \cdot \mbf{\Lambda} \cdot \mbf{L}(p),
\ee
where $\mbf{L}(p)$ is a \tit{standard boost}  taking the standard rest frame 4-momentum $\mbf{k}\equiv (m,0,0,0)$
to an arbitrary 4-momentum $\mbf{p}$, $\mbf{\Lambda}$ is an arbitrary LT taking
$\mbf{p}\to \mbf{\Lambda}\cdot \mbf{p} \equiv \mbf{\Lambda p}$, and
$\mbf{L}^{-1}(\Lambda p)$ is an inverse standard boost taking the final 4-momentum $\mbf{\Lambda p}$ back to
the particle's rest frame. Because of the form of the standard rest 4-momentum $\mbf{k}$, this final
rest momentum $\mbf{k}^{\sph '}$ can at most be a spatial rotation of the initial standard 4-momentum $\mbf{k}$, i.e.
$ \mbf{k}^{\sph '} = \mbf{W}(\Lambda,\vec{p}) \cdot \mbf{k}$. The rotation group $O(3)$ is then said to form
(Wigner's) \tit{little group} for massive particles, i.e. the invariance group of the particle's rest 4-momentum.
The explicit form of the standard boost is given by \cite{weinberg_qft1}
\bea{3}
%
 L^0_{\spp 0} &=& \gamma = \frac{p^{0}}{m}\no
 L^i_{\spp 0} &=& \frac{p^{i}}{m}, \quad L^0_{\spp i} = -\frac{p_{i}}{m}, \no
 L^i_{\sp j} &=& \delta^i_{\sp j} - (\gamma-1)\, \frac{p^i p_j}{|\vec{p}\,|^2}, \qquad i,j = (1,2,3),
\eea
where $\gamma = p^{0}/m = E/m \equiv e$ is the particles energy per unit rest mass. Note that
for the flat spacetime metric $\eta_{\a\b} =$diag$(1,-1,-1,-1)$, $p_0 = p^0$ and $p_i = -p^i$.

Peres, Scudo and Terno  \cite{peres_scudo_terno} considered a free spin $1/2$ particle in a normalizable state
containing a distribution of momentum states. Under a LT, \Eq{1} indicates that each component of the
momentum will undergo a different Wigner rotation, since the later is momentum dependent. Therefore, the
reduced spin density matrix for the particle, obtained  by tracing over the momentum states, will have
a non-zero von Neumann entropy, which will increase as with the rapidity $r$  of the
observer with constant velocity $v$ (with $\tanh r = v/c$). This indicates that the spin entropy of the
particle is not a relativistic scalar and thus has no invariant meaning.

Alsing and Milburn \cite{alsing_milburn} considered the transformation of Bell states of the form
\bea{4}
\ket{\Psi^\pm} &=& [\,\ket{\vec{p},\uparrow}\,\ket{-\vec{p},\uparrow} \pm
                      \ket{\vec{p},\downarrow}\,\ket{-\vec{p},\downarrow}\,]/\sqrt{2}, \no
\ket{\Phi^\pm} &=& [\,\ket{\vec{p},\uparrow}\,\ket{-\vec{p},\downarrow} \pm
                      \ket{\vec{p},\downarrow}\,\ket{-\vec{p},\uparrow}\,]/\sqrt{2},
\eea
composed of pure momentum eigenstates. Under a Lorentz boost perpendicular to the motion of
the particle, the transformed momentum $\vec{p}^{\sp '}$ is rotated by an angle $\theta$ with
respect to the original momentum $\vec{p}$, while the direction of the particle's spin
is rotated slightly less by the momentum dependent Wigner angle $\Omega_p < \theta$. The implication
is that if the boosted observer were to orient his detectors along $\pm\vec{p}^{\sp '}$ at angle $\pm\theta$ there could
be an apparent degradation of the violation of Bell inequalities in that inertial frame. However,
if the observer were to orient his detectors along the direction of the transformed spins $\pm\Omega_p$,
there would again be a maximal violation of the Bell inequalities for these states in the boosted inertial frame.
The entanglement of the complete bipartite state is preserved.

Gingrich and Adami \cite{adami_spin} considered a normalizable Bell state in which (analogous to Peres, Scudo and Terno)
there is a distribution over momentum states. Considering a Bell state $\ket{\Phi^+}$ with the momentum
in a product Gaussian distribution, a LT will transfer the spin entanglement into the momentum. If one forms the
reduced 2-qubit spin density matrix, it will exhibit a Wootter's concurrence \cite{wootters}
which decreases with increasing rapidity. Because the quantum states contain two degrees of freedom, spin and momentum,
the LT induces a \tit{spin-momentum entanglement} again due to the momentum dependent Wigner rotation \Eq{1}.
Similar investigations were carried out for the case of photons \cite{alsing_milburn,adami_photon,rembielinski}, which we
shall return to shortly.

\subsection{Curved Spacetime}
Subsequently, Terashima and Ueda \cite{ueda} extended the definition of the Wigner rotation of spin $1/2$ particles
to an arbitrary gravitational field. The essential point in going from SR to GR is that there are in general
no global inertial frames, only \tit{local frames}. In fact, in GR all reference frames are allowed, locally inertial
(i.e. zero acceleration trajectories - geodesics) or otherwise. Einstein's Equivalence Principle (EP) states
that in an arbitrary curved spacetime, SR holds locally in the tangent plane to a given event in spacetime
at the point $x$. The positive energy single particle states $\ket{\vec{p}(x),\sigma}$ now form the state space
for a \tit{local} representation of the Poincare group in the locally flat Lorentz tangent
plane at each spacetime point $x$.

Of particular importance is the observer's local reference frame which can be described by set of four axes
(4-vectors) called a tetrad $\bfe{a}(x)$ with $\ha\in\{0,1,2,3\}$ \cite{tetrads}. Three of these axes $\hi\in\{1,2,3\}$
describe the spatial axes at the origin of the observer's local laboratory, from which he makes local measurements.
The fourth axis $\ha=0$  describes the rate at which a clock, carried by the observer at the origin of his local laboratory,
ticks (gravitational redshift effect), and is taken to be the observer's 4-velocity $\bfe{0}(x)\equiv \mathbf{u}$,
where $\mathbf{u}$ is the tangent to the observer's worldline through spacetime.

In curved spacetime (CST) described by coordinates $x^\a$, we can define
\tit{coordinate basis vectors} (CBV) $\mbf{e}_\u(x)$ with components $e_\u^{\sp \a}(x) = \delta_\u^{\sp \a}$. We interpret the CBV $\mbf{e}_\u(x)$
as a local vector at the spacetime point $x$, pointing along the coordinate direction $x^\u$. These are
not unit vectors since their inner product yields the spacetime metric, $\mbf{e}_\u(x)\cdot\mbf{e}_\u(x) \equiv$
$g_{\a\b}(x)\,e_\u^{\sp \a}(x)\,e_\v^{\sp \b}(x) = g_{\u\v}(x)$. The observer's tetrad $\{\bfe{a}(x)\}$ form
an \tit{orthonormal basis}  (ONB, denoted by carets over the indices) such that the inner product of
any two basis vectors forms the flat spacetime metric of special relativity,
$\bfe{a}(x)\cdot\bfe{b}(x) \equiv$ $g_{\a\b}(x)\,\e{a}{\a}(x)\,\e{b}{\b}(x) = \eta_{\ha\hb}$
where $\eta_{\ha\hb}\equiv\,$ diagonal$(1,-1,-1,-1)$ is the metric of SR.
Note that $\e{a}{\a}(x)$ are the components of the ONB vectors written in terms of the CBVs,
$\bfe{a}(x)= \e{a}{\a}(x)\,\mathbf{e}_{\a}(x)$.
That the metric $g_{\a\b}(x)$
can be brought into the form $\eta_{\ha\hb}$ \tit{locally} at each spacetime point $x$  (since the
tetrad components $\e{a}{\a}(x)$ are spacetime dependent) is an embodiment of Einstein's Equivalence Principle (EP)
which states that SR holds locally at each spacetime point $x$ in an arbitrary curved spacetime (CST).
In Section~\ref{examples_CST} we will exhibit explicit examples of tetrads for various types of observers.

Tensor quantities $T^{\gamma\delta}_{\alpha\beta}(x)$
which ``live" in the surrounding spacetime described by coordinates $x^\a$, (called \tit{world} tensors, and
denoted by Greek indices), transform according to general coordinate transformations (GCT) $x^\a\to x'^\a(x)$,
which simply describe the same spacetime in the new coordinates. World tensors
can be projected down to  the observer's local frame via the components of the tetrad and its inverse,
i.e. $T^{\hc\hd}_{\ha\hb}(x) = \e{a}{\a}(x)\,\e{b}{\b}(x)\,\inve{\c}{c}(x)\,\inve{\d}{d}(x)\,T^{\gamma\delta}_{\alpha\beta}(x)$.
Here, $\inveT{a}{\u}(x)$ is the inverse transpose of the matrix of tetrad vectors $\e{a}{\u}(x)$, satisfying
$\mathbf{e}^{\ha}(x)\cdot\mathbf{e}^{\hb}(x) \equiv$ $g^{\a\b}(x)\,\inveT{a}{\a}(x)\,\inveT{b}{\b}(x) = \eta^{\ha\hb}$,
where $g^{\a\b}(x)$ and $\eta^{\ha\hb}$ are the inverses of the CST and flat spacetime metrics
$g_{\a\b}(x)$ and $\eta_{\ha\hb}$, respectively. We denote the observer's local components $T^{\hc\hd}_{\ha\hb}(x)$
of the world tensor $T^{\gamma\delta}_{\alpha\beta}(x)$ with hatted Latin indices. Objects with hatted
indices transform as scalars with respect to GCTs, but as \tit{local Lorentz vectors}
with respect to local Loretnz transformations (LLT)
$\Lambda(x)$, which simply transform between different instantaneous local Lorentz frames,
or instantaneous states of motion of different types of observers, at the point $x$
(e.g. stationary, freely falling, circular orbit, etc\ldots).

In particular, the local components $p^{\ha}(x)$  of the world 4-momentum $p^\a(x) = m\,u^\a(x)$
of a particle (with mass $m$ and 4-velocity $u^\a(x)$) passing through the observer's local laboratory
at the spacetime point $x$ are given by $p^{\ha}(x) = \inveT{a}{\a}(x)\,p^\a(x)= \mathbf{e}^{\ha}(x)\cdot\mathbf{p}(x)$.
In SR, $\mathbf{p}_{SR} = (E/c,\vec{p})$ is the 4-momentum of the particle, whose time component is
the particle's energy and whose spatial components are its 3-momentum.
In CST, $p^{\hat{0}}(x)$ is energy of the particle with 4-momentum $\mathbf{p}$ as measured by the observer
with tetrad $\{\mathbf{e}_{\ha}(x)\}$ as the particle passes through his local laboratory at the spacetime point $x$,
while $p^{\hi}(x)$ are the locally measured components of the 3-momentum.
Since GR allows for observer undergoing arbitrary motion (as opposed to SR which considers only zero acceleration or constant
velocity observers, i.e. inertial frames), the observer's locally measured components of the 4-momentum
$p^{\ha}(x)$ depends upon the observer's state of motion,
described by the motion of the axes comprising the his local laboratory, which are
given by  his tetrad $\{\mathbf{e}_{\ha}(x)\}$. As a field of vectors over the spacetime, we interpret the
the tetrad $\{\mathbf{e}_{\ha}(x)\}$ as a collection of observer's located at each spacetime point $x$,
usually of a particular type (stationary, freely falling, circular orbit, etc\ldots). We shall see explicit examples of different
types of observers (tetrads) in Section~\ref{examples_CST}.

As the passing (massive) particle moves in an arbitrary fashion from $x^\a\to x^{\a '} = x^\a + u^\a(x)\,d\tau$ in infinitesimal
proper time $d\tau$, Terashima and Ueda found that the local momentum components $p^{\ha}(x)$ would change under an
infinitesimal LT, $\lambda^{\ha}_{\spp \hb}(x)$ via $p^{\ha}(x)\to p^{\ha '}(x) = p^{\ha}(x) + \delta p^{\ha}(x)$ where
$\delta p^{\ha}(x) = \lambda^{\ha}_{\spp \hb}(x)\,p^{\hb}(x)$ and
$\Lambda^{\ha}_{\spp \hb}(x) = \delta^{\ha}_{\spp \hb}(x) + \lambda^{\ha}_{\spp \hb}(x) \, d\tau$
is the local Lorentz transformation (LLT) to first order in $d\tau$. A straight
forward, though lengthy calculation (detailed in Alsing \tit{et al} \cite{alsing_wigrot_spin}), leads from
$\lambda^{\ha}_{\spp \hb}(x)$ to the infinitesimal
Wigner rotation $\vartheta^{\ha}_{\spp \hb}(x)$ where
$W^{\ha}_{\spp \hb}(x) = \delta^{\ha}_{\spp \hb}(x) + \vartheta^{\ha}_{\spp \hb}(x)\,d\tau$ is the local Wigner rotation
to first order in $d\tau$. Terashima and Ueda showed that only the space-space components of $\vartheta^{\ha}_{\spp \hb}(x)$
are non-zero, and thus it truly represents an infinitesimal $\O(3)$ rotation for massive particles.

In the instantaneous non-rotating rest frame of the particle for geodesic motion (zero local acceleration),
Alsing \tit{et al} \cite{alsing_wigrot_spin} showed that the Wigner rotation would be zero. Further, considering the
$\O(\hbar)$ quantum correction to the non-geodesic motion of spin $1/2$ particles in an arbitrary gravitational field
the authors showed that in the instantaneous non-rotating rest frame of the accelerating particle (the
Ferimi-Walker (FW) frame) the Wigner rotation would also be observed to be zero.

\section{Photons}
\subsection{Flat Spacetime}
\label{CST_theory}
In the above, we have primarily considered massive spin $1/2$ particles. As briefly discussed above, analogous results
have been obtained for photons in SR \cite{alsing_milburn,adami_photon,rembielinski}. The fundamental difference between going from
massive to massless particles is there is no rest frame for the latter. Instead, one defines a \tit{standard frame}
in which the photon 4-momentum takes the form $\tilde{k}^\u=(1,0,0,1)$ where the photon travels in a
predefined direction along the $\mathbf{z}$-axis, and LT this to a photon of 4-momentum $\m{k}$ of
arbitrary energy $k^0$ and traveling in an arbitrary direction, subject to the null condition
$\m{k}\cdot\m{k}\equiv k^\u\,k_\u = 0$. The photon 4-momentum has the form
$\m{k} \leftrightarrow k^\u = (k^0,|\vec{k}|\,\hat{\mbf{k}})$ where $\vec{k}$ is the spatial 3-momentum of the
photon, $\hat{\mbf{k}} = \vec{k}/|\vec{k}| = \vec{n}$ is the direction of propagation of the photon and
$\omega\equiv k^0 = |\vec{k}|$ is the frequency of the photon.
Consequently, Wigner's little group
for massless particles is $SO(2)$, the group of rotations and translations in two dimensions associated
with the particle's transverse plane of polarization.
Under a LT $\Lambda$ taking $\m{k}\to\m{k}' = \Lambda\,\m{k}$  the transformation of the photon helicity state
$\ket{\m{k},\lambda}$, analogous to \Eq{1} is given by \cite{adami_photon,rembielinski}
\be{5}
U(\Lambda)\,\ket{\m{k},\lambda} = e^{i\,\lambda\,\psi(\Lambda,k)}\ket{\Lambda \m{k},\lambda},
\ee
where $\psi(\Lambda,k)$ is the momentum-dependent Wigner rotation angle (phase). Due to the fact
that $\m{k}$ is a null 4-vector,  The Wigner angle depends only on the
direction of propagation of the photon
$\hat{\mbf{k}}$ , and not on its frequency $\omega$, i.e.
$\psi(\Lambda,k) = \psi(\Lambda,\vec{n})$. Note that the unitary transformation in \Eq{5}
does not change the helicity of the photon state, in contrast to the case for massive particles \Eq{1}
in which the components of the spin are mixed up by a momentum dependent Wigner rotation.

The corresponding transformation of the polarization vectors  for positive and
negative helicity states $\epsilon^\mu_\pm$
\be{6}
\epsilon^\u_\pm(\hat{\mbf{k}}) = \frac{R(\hat{\mbf{k}})}{\sqrt{2}} \,
\left[
\begin{array}{c}
  0 \\
  1 \\
  \mp i \\
  0
\end{array}
\right],
\ee
is given by \cite{adami_photon}
\bea{7}
\epsilon_\pm^{'\u}(\hat{\mbf{k}'}) &\equiv& D(\Lambda)\,\epsilon_\pm^\u(\hat{\mbf{k}}) \no
&=& R(\Lambda \hat{\mbf{k}}) \, R_z(\psi(\Lambda,\vec{n})) \,
                          R(\hat{\mbf{k}})^{-1} \, \epsilon_\pm^\u(\hat{\mbf{k}}), \quad \no
&=& \Lambda  \epsilon_\pm^\mu(\hat{\mbf{k}}) - \frac{(\Lambda\,\epsilon_\pm^\mu(\hat{\mbf{k}}))^0}{(\Lambda\,k^\mu)^0} \, \Lambda  k^\mu.
\eea
where $R(\Lambda \hat{\mbf{k}})$ is the rotation taking the standard direction $\hat{\mathbf{z}}$ to $\hat{\mbf{k}}$.
Here we use the (abused) shorthand notation $\Lambda \hat{\mbf{k}} =\hat{\mbf{k}'}= \vec{k}'/|\vec{k}'|$.
In the the appendix we illustrate several examples in flat spacetime of cases where the Wigner angle is zero,
as well as cases in which it is non-zero, that will prove useful in our extension to curved spacetime below.

\subsection{Curved Spacetime}
In the following we extend the work of Terashima and Ueda \cite{ueda} and Alsing \tit{et al} \cite{alsing_wigrot_spin}
for massive spin $1/2$ particles in curved spacetime to photons.
Let $k_\u(x)$ be the photon 4-momentum, and $\e{a}{\u}(x)$ be the tetrad that defines
the (timelike) observer's local laboratory. The local (covariant) components of the photon 4-momentum $k_{\ha}(x)$,
measured in the observer's laboratory, are given by projecting $k_\u(x)$ onto the observer's local axes via the tetrad,
$k_{\ha}(x) = e_{\ha}^{\spp \u}(x)\,k_\u(x)$. We are interested in the change $\delta k_{\ha}(x)$ of the locally measured
photon components as the photon moves from $x^\u \to x^{'\u} = x^\u + k^\u(x)\,\d\xi$ (where $\xi$ is the
affine parameter along the photon's trajectory). Following \cite{ueda,alsing_wigrot_spin} we compute
\be{8}
\delta k_{\ha}(x) = \big(\delta e_{\ha}^{\spp \u}(x)\big)\,k_\u(x) + e_{\ha}^{\spp \u}(x)\,\delta k_\u(x).
\ee
For the last term we have
\be{9}
\delta k_\u(x) = d\xi\,\nabla_{\mathbf{k}}\,k_\u(x) \equiv
 d\xi\,k^\b(x) \nabla_{\b}\,k_\u(x)=0,
\ee
where $\nabla_{\b}$ is the Riemann covariant derivative \cite{tetrads} constructed from the spacetime metric.
The last equality is simply the definition that the photon's trajectory is a geodesic $\n{k}{\m{k}}=0$.

Using the orthonormality of the tetrad 4-vector axes $\inve{\v}{b}(x)\,\e{b}{\u}(x) = \delta_\v^{\spp \u}$,
the first term in \Eq{8} becomes
\bea{10}
\delta e_{\ha}^{\spp \u}(x) &=& d\xi\,\nabla_{\mathbf{k}}\, e_{\ha}^{\spp \u}(x) =
d\xi\,\big(\nabla_{\mathbf{k}}\, e_{\ha}^{\spp \v}(x) \big)\, \inve{\v}{b}(x)\,\e{b}{\u}(x), \no
&\equiv& \chi_{\ha}^{\spp \hb}(x)\,\e{b}{\u}(x)\,d\xi,
\eea
where we have defined the local matrix $\chi_{\ha}^{\spp \hb}(x)$ describing the rotation of the tetrad as
\be{11}
 \chi_{\ha}^{\spp \hb}(x) \equiv    \big(\nabla_{\mathbf{k}}\, e_{\ha}^{\spp \v}(x) \big)\, \inve{\v}{b}(x).
\ee
Thus, the change in the local components $k_{\ha}(x)$ of the photon's momentum as observed by the an observer with
tetrad $\e{a}{\u}(x)$ is given by the LLT
\bea{12}
\lefteqn{k_{\ha}(x) \to k'_{\ha}(x) \equiv k_{\ha}(x) + \delta k_{\ha}(x)}  \no
&=& \Lambda_{\ha}^{\spp \hb}(x)\, k_{\hb}(x) =
\left( \delta_{\ha}^{\spp \hb} + \lambda_{\ha}^{\spp \hb}(x)\,d\xi \right)\, k_{\hb}(x),
\eea
whose infinitesimal portion $\lambda_{\ha}^{\spp \hb}(x)$ is given by
\be{13}
 \delta k_{\ha}(x) = \chi_{\ha}^{\spp \hb}(x)\, k_{\hb}(x)\,d\xi \equiv \lambda_{\ha}^{\spp \hb}(x)\, k_{\hb}(x)\,d\xi.
\ee

Therefore, in CST the form of the transformation of a pure helicity photon state $\ket{\m{k},\lambda}$ has
the same form as \Eq{5} if we interpret $\m{k}$ as the the photon wavevector as measured by the
observer described by the tetrad $\bfe{a}(x)$, i.e. with components
$\m{k}\leftrightarrow k^{\ha}(x) = \inveT{a}{\a}(x)\, k^{\a}(x)$, and $\Lambda$ a LLT as given in \Eq{12}
for the infinitesimal motion of the photon along its geodesic from $x^\u \to x^{'\u} = x^\u + k^\u(x)\,\d\xi$.

\section{Wigner Rotation Angle for Photons in Curved Spacetime}
\subsection{Derivation}
Since in the instantaneous non-rotating rest frame of the observer the metric is locally flat
$\eta_{\ha\hb}=(1,-1,-1,-1)$ we can follow the SR derivation of the Wigner phase $\psi(\Lambda,k)$ by
Caban and Rembielinski \cite{rembielinski}, with the arbitrary LT in curved spacetime given by
$\Lambda^{\ha}_{\spp \hb}(x)$ in \Eq{12} to first order in $d\xi$. The authors' elegant derivation utilized the canonical
homomorphism between $SL(2,C)$ and the proper orthochronos homogeneous Lorentz group $L^{\uparrow}_+ \sim SO(1,3)$.
For every LLT $\Lambda(x)\in L^{\uparrow}_+$ of the photon 4-momentum $k^{\ha}(x)\to k^{'\ha}(x)\equiv\Lambda^{\ha}_{\spp \hb}(x)\,k^{\hb}(x)$
there corresponds a transformation of the Hermetian matrix $K(x)\equiv k^{\ha}(x)\sigma_{\ha}$ of the form
$K(x)\to K'(x) = A_{k'}(x) K(x) A^\dagger_k(x)$ where $A_{k}(x)\in SL(2,C)$, $\sigma_{\hi}$ are the constant
Pauli matrices and $\sigma_{\ho}$ is the $2\times 2$ identity matrix.
Henceforth we will drop the  spacetime argument $x$ on all local quantities for notational clarity.
For the most general photon 4-momentum $k^{\ha}=(k^{\ho},k^{\hi})$, $K$ has the form
\be{14}
K = k^{\ho} \,
\left(
\begin{array}{cc}
  1 + n^{\hat{3}} & n_- \\
  n_+& 1 - n^{\hat{3}}
\end{array}
\right),
\ee
where $n^{\hi} = k^{\hi}/k^{\ho}$ and $n_\pm = n^{\hat{1}} \pm i\, n^{\hat{2}}$.
The most general element $A\in SL(2,C)$ is of the form
\be{15}
A =
\left(
\begin{array}{cc}
  \alpha & \beta \\
  \gamma & \delta
\end{array}
\right),
\ee
where the entries are complex numbers subject only to the normalization condition $\trm{det}A = \a\d - \b\c=1$.
We wish to relate these entries to the components of infinitesimal LLT $\lambda^{\ha}_{\spp \hb}$, which is
antisymmetric $\lambda_{\hb\ha}=-\lambda_{\ha\hb}$.

Corresponding to the expression for the Wigner rotation $W(\Lambda,k) = L^{-1}_{k'} \Lambda L_k$ of \Eq{1}
with $p\to k$, is the  $SL(2,C)$ transformation
\be{16}
S(\Lambda,k) = A^{-1}_{k'} A A_k.
\ee
Here $A_k$ is the Lorentz boost that takes the standard photon 4-momentum
$\tilde{k}^{\ha}=(1,0,0,1)$ to $k^{\ha}=(k^{\hat{0}},k^{\hi})$, $A$ is an arbitrary LT \Eq{15}, and
$A^{-1}_{k'}$ is the inverse boost taking the transformed momentum $k'^{\ha}$ back to the standard 4-momentum $\tilde{k}^{\ha}$.
The most general form of $S(\Lambda,k)$, the $SL(2,C)$ element of Wigner's little group that leaves $\tilde{K}$
invariant,
is found by solving $\tilde{K} = S \tilde{K} S^\dagger_0$ yielding
\be{17}
S =
\left(
\begin{array}{cc}
  e^{i\psi/2} & z \\
  0 & e^{-i\psi/2}
\end{array}
\right),
\ee
where $\psi$ is the Wigner angle and $z$ is an arbitrary complex number.

Following \cite{rembielinski}, we now compute $S$ by the right hand expression in \Eq{16}, using the expression for $A$ in \Eq{15} for
an arbitrary LT. The $SL(2,C)$ element $A_k$ is obtained by solving $K = A_k \tilde{K} A^\dagger_k$
and is given by
\be{18}
A_k = \frac{1}{\sqrt{2(1+n^{\hat{3}})}} \,
\left(
\begin{array}{cc}
  1+n^{\hat{3}} & -n_- \\
  n_+ & 1+n^{\hat{3}}
\end{array}
\right) \,
\left(
\begin{array}{cc}
  \sqrt{k^{\ho}} &0 \\
  0 & 1/\sqrt{k^{\ho}}
\end{array}
\right),
\ee
corresponding to the product of a boost in the $z$-direction taking $\tilde{k}^{\ho}=1 \to k^{\ho}$, and
a rotation taking $\hat{\mbf{z}} \to \hat{\mbf{k}}$.

Lastly, we can compute the corresponding element of $A_{k'}\in SL(2,C)$ which transforms $K\to K'$ by an arbitrary
LT of the form \Eq{15}. Here $K'$ has the same form as \Eq{14} but with all quantities primed. The result is
\cite{rembielinski}
\bea{19}
K' &=& k'^{\ho} \,
\left(
\begin{array}{cc}
  1 + n'^{\hat{3}} & n'_- \\
  n'_+& 1 - n'^{\hat{3}}
\end{array}
\right) \no
&=& A_{k'}\,K\,A^\dagger_{k'} \equiv k^{\ho}\,
\left(
\begin{array}{cc}
  b & c^* \\
  c & a-b
\end{array}
\right),
\eea
where we have defined
\bea{20}
a &=& (|\a|^2 + |\c|^2)(1+n^{\hat{3}}) + (|\b|^2 + |\d|^2)(1-n^{\hat{3}}) \no
&+& (\a\b^*+\c\d^*) n_- + (\a^*\b+\c^*\d) n_+ = a^* \no
b &=& |\a|^2 (1+n^{\hat{3}}) + |\b|^2 (1-n^{\hat{3}}) + \a\b^* n_- + \a^*\b n_+ = b^* \no
c &=& \a^*\c (1+n^{\hat{3}}) + \b^*\d (1-n^{\hat{3}}) + \b^*\c n_- + \a^*\d n_+.\qquad
\eea
From \Eq{19} and \Eq{20} we can deduce the transformation of the photon 4-momentum as
\be{20.5}
k'^{\ho} = \frac{a}{2} k^{\ho}, \quad n'^{\hat{3}} = \frac{2 b}{a} -1, \quad n'_+ = \frac{2c}{a}
\ee
where the quantities $a,b,c$ in \Eq{20} depend only on the LT parameters $\a,\b,\c,\d$ of
$A_k$ and the \tit{direction} of the photon $n_{\pm},n^{\hat{3}}$, but not on the photon frequency $k^{\ho}$.

Forming the inverse of \Eq{19} and substituting it and \Eq{18} into the right hand side of \Eq{16}, and
subsequently equating the result to \Eq{17}, yields
\bea{21}
e^{i\psi(\Lambda,k)/2} &=& \frac{1}{a\sqrt{b(1+n^{\hat{3}})}} \, \left\{ [\a(1+n^{\hat{3}}) + \b n_+]b  \right.\no
&+& \left. [\c(1+n^{\hat{3}}) + \d n_+] c^* \right\}.
\eea
Note that the expression for the arbitrary complex number $z$ is not needed since it does not enter into
the expression for the transformation of the photon helicity state \Eq{5}.

Finally, to relate the entries of \Eq{15} to the infinitesimal LLT $\lambda^{\ha}_{\spp \hb}$, we
expand $K' = A K A^\dagger$ in terms of $k^{\ha}$ and
$k^{'\ha}\equiv\Lambda^{\ha}_{\spp \hb}\,k^{\hb}$ and expand $\Lambda^{\ha}_{\spp \hb}$ as in \Eq{12}.
Multiplying by a general Pauli matrix and using the relation $\half tr(\sigma_{\ha}\sigma_{\hb}) = \delta_{\ha\hb}$
yields the expression
\be{22}
\lambda^{\ha}_{\spp \hb} = \half \eta^{\ha\hc}\, tr\left(\sigma_{\hb}\sigma_{\hc} A
                         + \sigma_{\hc}\sigma_{\hb} A^\dagger \right),
\ee
where $tr$ is the matrix trace.
Since we are interested in an infinitesimal LLT
we further expand $A$  as
\be{23}
A =
\left(
\begin{array}{cc}
  \alpha & \beta \\
  \gamma & \delta
\end{array}
\right) \equiv
I + \tilde{A}\,d\xi, \quad
\tilde{A} =
\left(
\begin{array}{cc}
  \tilde{\alpha} & \tilde{\beta} \\
  \tilde{\gamma} & -\tilde{\alpha}
\end{array}
\right),
\ee
which satisfies $\det(A)=1$  to first order in $d\xi$.
A straightforward calculation leads to
\bea{24}
\tilde{\a} &=& \half \,\left( \lambda^{\hat{0}}_{\spp \hat{3}} - i \lambda^{\hat{1}}_{\spp \hat{2}} \right), \no
\tilde{\b} &=& \half \,\left[ \left( \lambda^{\hat{0}}_{\spp \hat{1}} + \lambda^{\hat{3}}_{\spp \hat{1}} \right)
-i \left( \lambda^{\hat{0}}_{\spp \hat{2}} + \lambda^{\hat{2}}_{\spp \hat{3}} \right) \right]  \no
\tilde{\c} &=& \half \,\left[ \left( \lambda^{\hat{0}}_{\spp \hat{1}} - \lambda^{\hat{3}}_{\spp \hat{1}} \right)
+i \left( \lambda^{\hat{0}}_{\spp \hat{2}} - \lambda^{\hat{2}}_{\spp \hat{3}} \right) \right],
\eea
with $\lambda^{\ha}_{\spp \hb}(x) = \chi^{\ha}_{\spp \hb}(x)$ given in terms of the observer's tetrad by \Eq{11}.
Using the above expressions for $\tilde{A}$ we can form the entries of $A$, and using the expression for
$a,b,c$ from \Eq{20} determine the infinitesimal Wigner rotation angle $\tilde{\psi}(\Lambda,\vec{n})$
when we expand \Eq{21} to $\O(d\xi)$ as
\be{25}
e^{i\psi(\Lambda,k)/2} \sim 1 + i\tilde{\psi}(\Lambda,k)\,d\xi/2.
\ee
Finite Wigner rotations can be built up as a time ordered integration of infinitesimal Wigner rotations
over the geodesic trajectory $x(\xi)$  of the photon (obtained by solving \Eq{9}) via
\be{26}
\exp\Big(i\psi(\Lambda,\vec{n})/2\Big) = T\exp\left[ i\int  \tilde{\psi}\Big(\Lambda,\vec{n}(\xi)\Big)\,d\xi/2 \right]
\ee
where $\vec{n}(\xi) = \vec{n}(x(\xi))$, $\Lambda^\u_{\spp \v}(\xi) = \Lambda^\u_{\spp \v}(x(\xi))$ and
$T$ is the time order operator.

\subsection{Examples of the Wigner rotation angle in the Schwarzschild metric}
\label{examples_CST}
We now consider some specific examples of the Wigner rotation angle $\psi(\Lambda,k)$ as computed
from \Eq{20} and \Eq{21} in the static, spherically symmetric Schwarzschild spacetime,
and compare and contrast the results with SR, which holds locally at each spacetime point.
The Schwarzschild metric is given by
\be{27}
ds^2 = (1 - r_s/r)\,c^2dt^2 - \frac{dr^2}{(1 - r_s/r)} + r^2\,(d\theta^2 + r^2\,\sin^2\theta\,d\phi^2)
\ee
where $r_s = 2GM/c^2$ is the Schwarzschild radius of the central gravitating object of mass $M$
(e.g. for the Earth $r_{s\oplus} = 0.89$ cm, and for the Sun $r_{s\odot} = 2.96$ km).
Henceforth, we use units where $G=c=1$. Since the metric is independent of $\phi$,
orbital angular momentum is conserved \cite{note_1}, so without
loss of generality we consider photon orbits in the equatorial plane $\theta=\pi/2$.

\subsubsection{Radially infalling photons}
The 4-momentum for a radially in-falling photon (satisfying $\nabla_{\mathbf{k}}\, \mathbf{k}=0$)
with zero orbital angular momentum is given by
\bea{28}
k^\u(x) &\equiv& \left(k^t(x),\, k^r(x),\, k^\theta(x),\, k^\phi(x)\,\right) \no
        &=& \omega\,\big( 1/(1-r_s/r),\, -1, 0,\, 0\, \big),
\eea
satisfying $\mathbf{k}\cdot\mathbf{k}\equiv g_{\u\v}\,k^\u\,k^\v=0$.
The constant (of the motion) $\omega$ is the frequency of the photon as measured by a stationary observer (discussed below)
at spatial infinity ($r\to\infty$).

We now consider several different types of (massive) observers defined by stating their associated tetrads.
An obvious first choice is to consider a \tit{stationary} observer who sits at fixed spatial coordinates
$(r,\, \theta,\, \phi\,)$ whose tetrad is given by
\bea{29}
(e^{stat}_{\hat{0}})^\u(x) &=& \left( 1/(1-r_s/r)^{1/2},0,0,0\right)=\mathbf{e}^{stat}_{\hat{t}}, \no
(e^{stat}_{\hat{3}})^\u(x) &=& \left( 0,(1-r_s/r)^{1/2},0,0\right)=\mathbf{e}^{stat}_{\hat{r}}, \no
(e^{stat}_{\hat{1}})^\u(x) &=& \left( 0,0,1/r,0\right)=\mathbf{e}^{stat}_{\hat{\theta}}, \no
(e^{stat}_{\hat{2}})^\u(x) &=& \left( 0,0,0,1/r\right)=\mathbf{e}^{stat}_{\hat{\phi}},
\eea
Since we are considering motion in the equatorial plane, we have oriented our coordinate system so that
the local $\hat{3}$ axis points along the (increasing) radial direction, the $\hat{2}$ axis points along the
(increasing) $\phi$ direction, then the $\hat{1}$ axis direction (normal to the equatorial plane) points
in the (increasing) $\theta$ direction. Note that the world components of the tetrad vectors $\e{a}{\a}(x)$
are given in the order $\alpha = (0,1,2,3) \leftrightarrow (t,r,\theta,\phi)$, while
the ordering of our observer's local axes are given in the order $\ha=(\ho,\hat{1},\hat{2},\hat{3})\leftrightarrow$
$(\hat{t},\hat{\theta},\hat{\phi},\hat{r})$ corresponding to the observer's local time and
$\hat{\mathbf{x}}, \hat{\mathbf{y}}, \hat{\mathbf{z}}$ axes.

A stationary observer must exert an acceleration $\mathbf{a}$ defined by $\mathbf{a} = \nabla_{\mathbf{u}}\,\mathbf{u}$
in order to oppose the gravitational attraction of the central mass $M$ and remain at a fixed spatial location.
(Note that in SR a stationary observer undergoes zero acceleration since there is no gravitational field, i.e. $M=0$).
In general, the magnitude of the local acceleration experienced by the observer in
his local frame is $a\equiv ||\mathbf{a}|| \equiv (-\mathbf{a}\cdot\mathbf{a})^{1/2}$ (where the minus sign results
from the fact that $\mathbf{a}$ is a spacelike vector). For the stationary observer, the local acceleration
is given by $a^{stat}=(M/r^2)\,(1-r_s/r)^{-1/2}$, which approaches the usual Newtonian
form of $M/r^2$ as $r\to\infty$. The fact that
$\lim_{r\to r_s} a^{stat} = \infty$ indicates that for $r<r_s$, i.e. inside the
 event horizon of the black hole, the observer can no longer remain stationary and is inexorably drawn into
 the singularity of the black hole.

From the above tetrad we can compute the local components $k^{\ha}(x)$ of the photon's 4-momentum $k^\u(x)$
as measured in the observer's local frame with $k^{\ha}(x) = \inveT{a}{\u}(x)\,k^\u(x)$. Here,
$\inveT{a}{\u}(x)$ is the inverse transpose of the matrix of tetrad vectors $\e{a}{\u}(x)$.
The unit vector $n^{\hi}$, used in \Eq{20} and \Eq{21},
which describes the direction of the photon as measured in the observer's local frame is given by
$n^{\hi} = k^{\hi}/(k^{\hi}\,k_{\hi})^{1/2}$
where $k^{\hi}\,k_{\hi} = \sum_{\hi=\hat{1}}^{\hat{3}} \, (k^{\hi})^2$ is the ordinary Euclidean flat spacetime  dot product of the
spatial 3-vector  portion $k^{\hi}$ of the local 4-vector $k^{\ha}$. For the radially infalling photon that we
consider, we have $n^{\hi}= (0,\, 0,\, -1\,)$.

The general formula for the Wigner rotation angle $\psi(\Lambda,\vec{n})$ is given in \Eq{21}. Since
all relevant quantities in this formula are a function of the spacetime point $x$, we are interested
in computing the infinitesimal Wigner rotation angle  $\tilde{\psi}(\Lambda,k)$ as defined in \Eq{25}.
This requires the infinitesimal version of the entries $\{\a, \b, \c, \d\}$ of the $SL(2,C)$ matrix $A$ in \Eq{23}
which describes the LLT. We defined these infinitesimal entries as $\{\tilde{\a}, \tilde{\b}, \tilde{\c}, \tilde{\d}\}$
through the $SL(2,C)$ matrix $\tilde{A}$ in the same equation, which described the infinitesimal LLT.
From \Eq{24}, $\{\tilde{\a}, \tilde{\b}, \tilde{\c}, \tilde{\d}\}$ are related to
the $4\times 4$ infinitesimal LLT $\lambda^{\ha}_{\spp \hb}$, which from \Eq{13} is equal to the
matrix $\chi^{\ha}_{\spp \hb}$ describing the rotation of the tetrads. Finally, $\chi^{\ha}_{\spp \hb}$
can be computed from \Eq{11} given the observer's tetrad and the photon 4-momentum.
Thus, combining all these infinitesimal contributions into the right hand side of \Eq{21} and equating
this to \Eq{25} allows us to extract the Wigner rotation angle $\psi(\Lambda,\vec{n})$ to first order in $d\xi$.

Applying the above prescription to the radially infalling photon and the stationary observer leads to
the result $\psi(\Lambda,\vec{n})=0$. This is an expected
result since as the photon moves along its radial geodesic it is continually boosted in the same direction.
As in the flat spacetime case, an stationary observer in the path of the photon detects no Wigner rotation
of the photon's polarization \cite{rembielinski}.

As discussed above, the stationary observer in a static metric undergoes a non-zero acceleration to
remain at a fixed spatial location. The general relativistic observer that is locally analogous to the inertial (constant
velocity) observer of SR is the \tit{freely falling frame} (FFF).
The FFF is defined by the tetrad $\bfe{a}(x)$ satisfying the condition
\be{30}
\nabla_{\mathbf{u}}\, \bfe{a} = 0.
\ee
Since $\bfe{0}(x) = \mathbf{u}(x)$ is the 4-velocity of the observer's geodesic, the  $\ha = \hat{0}$ equation
in \Eq{30} is just the geodesic equation, stating that the FFF observer experience zero local acceleration
$||\mathbf{a}^{FFF}||=0$.
The remaining $\ha = \hi$ equations state that the spatial tetrad axes
$\bfe{i}(x)$ are parallel transported along the observer's geodesic.

In the Schwarzschild metric, the tetrad for a radially FFF observer is given by
\bea{31}
(e^{FFF}_{\hat{0}})^\u(x) &=& \left( (1-r_s/r)^{-1},-(r_s/r)^{1/2},0,0\right)=\mathbf{e}^{FFF}_{\hat{t}}, \no
(e^{FFF}_{\hat{3}})^\u(x) &=& \left( -(r_s/r)^{1/2}\,(1-r_s/r)^{-1},1,0,0\right)=\mathbf{e}^{FFF}_{\hat{r}}, \no
(e^{FFF}_{\hat{1}})^\u(x) &=& \left( 0,0,1/r,0\right)=\mathbf{e}^{FFF}_{\hat{\theta}}, \no
(e^{FFF}_{\hat{2}})^\u(x) &=& \left( 0,0,0,1/r\right)=\mathbf{e}^{FFF}_{\hat{\phi}},
\eea
Following the above prescription with the FFF tetrad, we again find that $\psi(\Lambda,\vec{n})=0$.
Since SR holds at each spacetime point, we could have invoked Einstien's equivalence principle to
deduce this last result.

\subsubsection{Photon with nonzero angular momentum}
A general photon orbit in the Shcwarzschild metric obeys the radial ``energy" equation \cite{hartle}
\be{32}
\frac{1}{b_{ph}^2} = \left( \frac{dr}{d\xi} \right)^2 + W_{eff}(r), \quad
W_{eff}(r) = \frac{1}{r^2}\,\left(1-\frac{r_s}{r}\right),
\ee
where $\xi$ is the affine parameter along the photon geodesic, and $k^\u = dx^\u/d\xi$.
In \Eq{32}, the quantity $1/b^2_{ph}$ acts as an effective energy. Here $b_{ph}=|l_{ph}/e_{ph}|$ is the ratio of
the orbital angular momentum $l_{ph}$ and the energy $e_{ph}$ of the photon, both of which are constant since
the Schwarzschild metric is independent of $t$ and $\phi$, respectively. For $r\gg r_s$ the
quantity $b$ has the interpretation of the impact parameter of the photon with respect to $M$ situated
at $r=0$. The most general infalling photon  geodesic (starting at spatial infinity) in the equatorial plane ($k^\theta=d\theta/d\xi=0$ ) is given by
\be{33}
k^\u(x) = \left(\frac{1}{(1-r_s/r)}, -\left(1-\frac{b_{ph}^2(1-r_s/r)}{r^2}\right)^{1/2}, 0, \frac{b_{ph}}{r^2} \right),
\ee
which reduces to the radial infalling photon in the equatorial plane \Eq{28}, for $b_{ph}\to 0$.
Unlike the flat spacetime case of SR, the orbit of photon in \Eq{32} is curved, and there even exits
an unstable circular orbit for $b^2_{ph} = 27 M^2$ for which $W_{eff}$ has a maximum value.
For stationary metrics, one can interpret the the bending of light in an optical-mechanical analogy
in which the metric can be viewed as a spatially varying index of refraction,
that takes on a value of unity at spatial infinity and is infinite at the event horizon \cite{grav_n}.
For $1/b^2_{ph} < 1/ (27 M^2)$ there is a turning point in the photon orbit, and the photon will again
escape to infinity. For $1/b^2_{ph} > 1/(27 M^2)$ the photon spirals into the event horizon and is captured
by the black hole.

An observer (massive) traveling on a geodesic in the equatorial plane satisfies the corresponding
radial ``energy" equation \cite{hartle}
\bea{34}
{\mathcal{E}}\equiv \frac{e^2_{obs}-1}{2} &=& \half\left( \frac{dr}{d\tau} \right)^2 + V_{eff}(r), \no
V_{eff}(r) &=& -\frac{M}{r} + -\frac{l_{obs}}{2 r^2} -\frac{M l^2_{obs}}{r^3},
\eea
where the constants of the motion $e_{obs} = (1-r_s/r)\,dt/d\tau$ and $l_{obs} = r^2 sin^2\theta\,d\phi/d\tau$
(here $\theta=\pi/2$)  can be interpreted, at large $r$, as the observer's energy per unit mass and
orbital angular momentum per unit mass. Here $\tau$ is the observer's proper time defined from the metric  \Eq{27} as
$c\,d\tau = ds$.
In general, the FFF satisfying $\nabla_{\mathbf{u}}\, \bfe{a} = 0$ depends separately on both $e_{obs}$ and
$l_{obs}$. To keep the algebra manageable, in the following we will consider the observer's geodesic to
lie in the equatorial plane ($u^\theta = d\theta/d\tau=0$ with $e_{obs}=1$ but arbitrary orbital
angular momentum $l_{obs}$. This FFF tetrad, which we denote by $\bfe{a}^{FFF(l)}(x)$ is given by
\begin{widetext}
\bea{35}
(e^{FFF(l)}_{\hat{0}})^\u(x) &=& \left( \frac{1}{(1-r_s/r)},u^r,0,\frac{l_{obs}}{r^2}\right)=\mathbf{e}^{FFF(l)}_{\hat{t}}, \no
(e^{FFF(l)}_{\hat{3}})^\u(x) &=&
\left(
-\sqrt{\frac{r_s}{r}}\,\frac{\cos\Phi(r)}{(1-r_s/r)},
-\sqrt{\frac{r_s}{r}}\,u^r\,\cos\Phi(r) - \frac{l_{obs}\,(1-r_s/r)\,\sin\Phi(r)}{\sqrt{r_s r}},
0,
\frac{l_{obs}\,\cos\Phi(r)}{\sqrt{r_s r^3}} - \frac{u^r\,\sin\Phi(r)}{\sqrt{r_s r}}
\right)=\mathbf{e}^{FFF(l)}_{\hat{r}}, \no
(e^{FFF(l)}_{\hat{1}})^\u(x) &=& \left( 0,0,1/r,0\right)=\mathbf{e}^{FFF(l)}_{\hat{\theta}}, \no
(e^{FFF(l)}_{\hat{2}})^\u(x) &=&
\left(
-\sqrt{\frac{r_s}{r}}\,\frac{\sin\Phi(r)}{(1-r_s/r)},
\frac{l_{obs}\,(1-r_s/r)\,\cos\Phi(r)}{\sqrt{r_s r}} + \sqrt{\frac{r}{r_s}}\,u^r\,\sin\Phi(r) ,
0,
- \frac{u^r\,\cos\Phi(r)}{\sqrt{r_s r}} + \frac{l_{obs}\,\sin\Phi(r)}{\sqrt{r_s r^3}}
\right)=\mathbf{e}^{FFF(l)}_{\hat{\phi}}, \no
\eea
\end{widetext}
where the radial component of the observer's 4-velocity $u^r = dr/d\tau$ and
the angle of rotation $\Phi(r)$ of the spatial tetrads in the equatorial plane are
given by
\bea{36}
u^r(r) &=& -\left(\frac{r_s}{r} - \frac{l^2_{obs}}{r^2}\,(1-r_s/r) \right)^{1/2}, \no
\frac{d\Phi(r)}{dr} &=& -\frac{l_{obs}}{2\,r^2\,u^r(r)}.
\eea

The photon and observer geodesics given by \Eq{33} and \Eq{35}, respectively, both lie in the
equatorial plane ($\theta = \pi/2$).
From the discussion in the appendix of the Wigner rotation in flat spacetime,
we can invoke the EP to associate the observer's local (spatial tetrad) axes $(\hat{1},\hat{2},\hat{3})$
with the inertial axes $\hat{\mathbf{x}}, \hat{\mathbf{y}}, \hat{\mathbf{z}}$ used in the appendix to discuss
the Wigner rotation in the flat spacetime of SR. For both the photon and observer geodesics in the
equatorial plane $\hat{2}$-$\hat{3}$ ($\bfe{\phi}\,$-$\,\bfe{r}$) corresponding to the SR
$\hat{\mathbf{y}}$-$\hat{\mathbf{z}}$ plane used in the appendix, the Wigner rotation is identically zero,
$\psi(\Lambda,\vec{n})=0$ (see the discussion in the appendix following \Fig{figA1}). This is borne out
by a GR calculation utilizing \Eq{33} and \Eq{35}.

Following the appendix, to obtain a non-zero Wigner angle in flat spacetime, we considered
the situation with the photon moving in the $\hat{\mathbf{x}}$-$\hat{\mathbf{y}}$ with
the observer moving in the $\hat{\mathbf{y}}$-$\hat{\mathbf{z}}$ plane (see the discussion
in the appendix of \Fig{figA3}). Invoking the EP this corresponds in the GR case to the
photon's geodesic being in the $\hat{1}$-$\hat{2}$ or $\bfe{\theta}\,$-$\,\bfe{\phi}$ plane, while
the observer's geodesic remains in the equatorial plane, $\hat{2}$-$\hat{3}$, or $\bfe{\phi}\,$-$\,\bfe{r}$ plane.
We can modify the photon geodesic lying in the equatorial plane $\theta=\pi/2$ \Eq{33},
to a geodesic lying in the plane $\phi = \pi/2$
with the expression
\bea{37}
k^\u(x) &=& \left(\frac{1}{(1-r_s/r)}, -\left(1-\frac{b_{ph}^2(1-r_s/r)}{r^2}\right)^{1/2}, \frac{b_{ph}}{r^2},0 \right), \no
&\approx& \left( \frac{1}{(1-r_s/r)}, -1, \frac{b_{ph}}{r^2}, 0 \right) + O(b^2_{ph}).
\eea
In the last line of \Eq{37} we have expanded the photon 4-vector to first order in $b_{ph}$, to keep the
algebra manageable in the following calculation. Similarly, expanding \Eq{35} to first order in $l_{obs}$
we obtain
\bea{38}
(e^{FFF(l)}_{\hat{0}})^\u(x) &=& \left( \frac{1}{(1-r_s/r)},-\sqrt{\frac{r_s}{r}},0,\frac{l_{obs}}{r^2}\right)=\mathbf{e}^{FFF}_{\hat{t}}, \no
(e^{FFF(l)}_{\hat{3}})^\u(x) &=& \left( \sqrt{\frac{r_s}{r}}\,\frac{1}{(1-r_s/r)},1,0,-\frac{2 l_{obs}}{\sqrt{r_s r^3}}\right)=\mathbf{e}^{FFF}_{\hat{r}}, \no
(e^{FFF(l)}_{\hat{1}})^\u(x) &=& \left( 0,0,1/r,0\right)=\mathbf{e}^{FFF}_{\hat{\theta}}, \no
(e^{FFF(l)}_{\hat{2}})^\u(x) &=& \left( \frac{-l_{obs}}{r}\,\frac{1}{(1-r_s/r)},\frac{l_{obs}\,(2-r_s/r)}{\sqrt{r_s r}},0,\frac{1}{r}\right)\no
&=&\mathbf{e}^{FFF}_{\hat{\phi}},
\eea
which satisfies the FFF tetrad conditions \Eq{30} and
$\mathbf{e}\cdot \mathbf{g} \cdot \mathbf{e}^T = \mbf{\eta}$ to correction terms of  $O(l^2_{obs})$.
Computing $\tilde{\psi}(\Lambda,k)$ in \Eq{25} we find a non-zero (infinitesimal) Wigner rotation angle.
\be{39}
\tilde{\psi}(\Lambda,k) = \frac{b_{ph}\,l_{obs}}{r^3}\,
\left(
1 + \frac{3}{2}\, \sqrt{\frac{r}{r_s}} - \frac{5}{4}\, \sqrt{\frac{r_s}{r}}.
\right)
\ee
Note that the above result is proportional to both $b_{ph}$ and $l_{obs}$. If $b_{ph}=0$, the photon would
be radial in the plane $\phi=\pi/2$ and would therefore only intersect the observer's curved geodesic in the
equatorial plane $\theta=\pi/2$ at $r=0$.
Similarly, if $l_{obs}=0$, the observer's geodesic would be radial in the equatorial plane, and intersect
the photons curved geodesic again at $r=0$.
\begin{figure}[h]
\includegraphics[width=3.75in,height=2.75in]{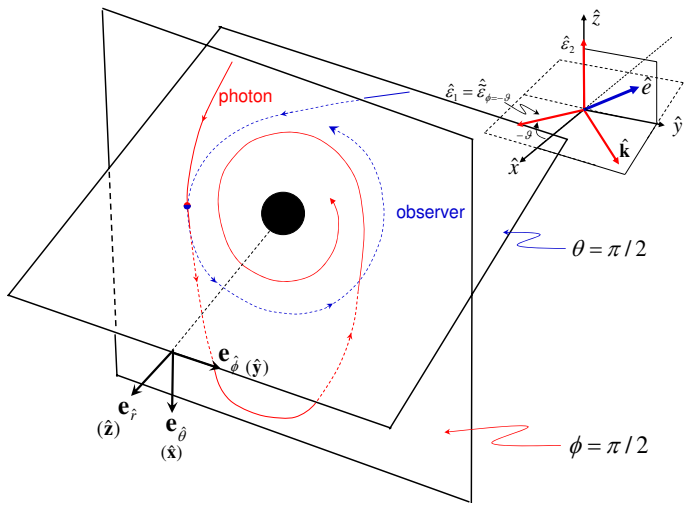}
\caption{(Color online) Example of a nonzero Wigner angle measured by an observer in Schwarzschild spacetime.
In Schwarzschild coordinates $x^\a=(t,r,\theta,\phi)$,
the photon geodesic lies in the plane $\phi=\pi/2$, while the observer's geodesic lies in the $\theta=\pi/2$
equatorial plane. The photon geodesic has a nonzero impact parameter $b_{ph}$, while the
observer's geodesic has nonzero angular momentum $l_{obs}$. The two-toned colored circle shows
the intersection of photon and observer geodesics. Compare with the SR flat spacetime case
illustrated in \Fig{figA3}, which by the EP, holds at the intersection point.}
\label{fig1}
\end{figure}
For both  $b_{ph}$ and $l_{obs}$ non-zero we have the trajectories as illustrated in \Fig{fig1}.
Note that in general the photon and and observer's geodesic intersect at one spacetime point
(the two-toned circle in the figure). The observer at this instant in time and this location,
observing the photon traversing his local laboratory, will measure the nonzero infinitesimal
Wigner angle in \Eq{39}.

\section{General Frames in which the Wigner rotation angle is zero}
\subsection{Conditions}
We now explore the condition under which the observer will observe zero Wigner rotation of the photon's polarization.
From \Eq{13} the change $\delta k_{\ha}(x)$ in the local components of the photon wavevector
is zero if $ \lambda_{\ha}^{\spp \hb}(x) =  \chi_{\ha}^{\spp \hb}(x) =0$. From \Eq{11} a sufficient condition for this
to occur is
\be{26psi0}
\nabla_{\mathbf{k}}\, e_{\ha}^{\spp \v}(x) = 0.
\ee
In the case of massive particles considered by Alsing \tit{et al} \cite{alsing_wigrot_spin}, the photon
momentum $\mathbf{k}$ would be replaced by the massive particle's 4-velocity $\mathbf{u} = \bfe{0}$.
The corresponding condition $\nabla_{\mathbf{u}}\, e_{\ha}^{\spp \v}(x) = 0$ defines the instantaneous
non-rotating rest frame of the particle traveling on a geodesic (zero acceleration - if one
ignores the particle's spin), i.e. the observer's local laboratory rides along with
the passing particle. Since there is no rest frame for a photon, \Eq{26psi0} describes something different,
and it is not immediately obviously that a solution for the observer's local laboratory, described
by the tetrad $e_{\ha}^{\spp \v}(x)$, exists.

We have found solutions to \Eq{26psi0}, which describe a class of observers situated at each spacetime point $x$,
which are in fact Fermi-Walker frames (FWF). The FWF is the instantaneous non-rotating rest frame of
a particle experiencing arbitrary non-zero acceleration $\mathbf{a} = \nabla_{\mathbf{u}}\,\mathbf{u}$, and
is defined by the following equation
\be{27psi0}
\nabla_{\mathbf{u}}\, \mathbf{s} = ( \mathbf{u}\cdot\mathbf{s} ) \,
                      \mathbf{a} - ( \mathbf{a}\cdot\mathbf{s} ) \, \mathbf{u}.
\ee
A vector $\mathbf{s}$ satisfying \Eq{27psi0} is said to be \tit{Fermi Walker transported}.
If $\mathbf{s}=\mathbf{u}$, \Eq{27psi0} reproduces the definition of the acceleration $\mathbf{a}$.
In general, $\mathbf{s}$ is any one of the three orthonormal
spatial axes $e_{\ha}^{\spp \hi}(x)$ of the observer's tetrad that is orthogonal to $\mathbf{u}$,
and \Eq{27psi0} is a generalization of parallel transport when $\mathbf{a}\ne 0$. Equation~(\ref{27psi0})
defines a locally \tit{non-rotating} frame in the sense that if $\mathbf{s}$ is orthogonal to the
instantaneous osculating plane defined by $\mathbf{u}$ and $\mathbf{a}$ then
$\nabla_{\mathbf{u}}\, \mathbf{s}=0$, i.e. $\mathbf{s}$ is parallel transported along the world line with
tangent $\mathbf{u}$.


\subsection{Sufficient condition for existence of zero Wigner rotation angle frames}
That solutions of \Eq{26psi0} can be found that also satisfy \Eq{27psi0} is not readily obvious since the
later FW transport equation is a statement solely about the observer and therefore is independent
of the photon 4-momentum $\mathbf{k}$. To show that solutions of \Eq{26psi0} are compatible with \Eq{27psi0}
let $\mathbf{s}$ be any of the three orthonormal spatial vectors of the tetrad, and $\mathbf{u}=\bfe{t}$
be the observer's 4-velocity. Hence, $\mathbf{u}\cdot\mathbf{u}=1$,  $\mathbf{s}\cdot\mathbf{s}=-1$,
and $\mathbf{u}\cdot\mathbf{s}=0$. Equation~(\ref{26}) is then restated as
\be{34psi0}
\nabla_{\mathbf{k}}\e{a}{\v}(x)=0, \quad \Rightarrow \quad
\nabla_{\mathbf{k}}\mathbf{u}=0, \; \nabla_{\mathbf{k}}\mathbf{s}=0.
\ee
As a integrability condition, we take $\nabla_{\mathbf{k}}$ of the FW transport equation \Eq{27psi0}
and use \Eq{34psi0} repeatedly along with the orthogonality of $\mathbf{u}$ and $\mathbf{s}$,
noting that $\mathbf{a} = \nabla_{\mathbf{u}} \mathbf{u}$. This yields the equation
\be{35psi0}
\nabla_{\m{k}}\,\nabla_{\m{u}}\, \m{s} + [(\nabla_{\m{k}}\nabla_{\m{u}}\m{u} )\cdot\m{s}] \, \m{u} = 0.
\ee
Projecting \Eq{35psi0} onto the the particle's 4-velocity by taking its dot product with $\m{u}$ yields
\be{36psi0}
( \n{k}\,\n{u} \, \m{s} ) \cdot \m{u} + ( \n{k}\,\n{u} \, \m{u} ) \cdot \m{s} = 0.
\ee
Equation~(\ref{36psi0}) can be obtained independently by differentiating the orthogonality condition $\m{s}\cdot\m{u}=0$,
first with respect to $\n{u}$ and then with respect to $\n{k}$ and making repeated use of \Eq{34psi0}.

Similarly, by projecting \Eq{35psi0} onto an arbitrary spatial vector $\m{s}'$ of the tetrad we obtain
\be{37psi0}
\big(\nabla_{\m{k}}\,\nabla_{\m{u}}\, \m{s} \big)\cdot \m{s}'= 0.
\ee
Since we could have equivalently written down \Eq{35psi0} using $\m{s}'$ and then subsequently projected onto $\m{s}$,
\Eq{37psi0} must also hold with these two vectors reversed, i.e.
\be{38psi0}
\big(\nabla_{\m{k}}\,\nabla_{\m{u}}\, \m{s}' \big)\cdot \m{s}= 0.
\ee
Since $\m{s}'\cdot\m{s}=-1$ if $\m{s}'=\m{s}$, and zero otherwise, applying first $\n{u}$ and then
$\n{k}$ to the relation $\m{s}'\cdot\m{s}=\trm{constant}$, and again repeatedly using \Eq{34psi0},
yields
\be{39psi0}
\big(\nabla_{\m{k}}\,\nabla_{\m{u}}\, \m{s} \big)\cdot \m{s}'
+ \big(\nabla_{\m{k}}\,\nabla_{\m{u}}\, \m{s}' \big)\cdot \m{s}= 0.
\ee
One possible solution of \Eq{39psi0} is that each separate term is identically zero as in \Eq{37psi0} and \Eq{38psi0}.
This shows that a solution of  \Eq{26psi0} has the FW transport equation \Eq{27psi0} as a sufficient condition,
though not a necessary condition.  For example, in Schwarzschild spacetime we can explicitly solve for the
tetrad for a radially accelerating FWF observer for case of a radially infalling photon,
both in the equatorial plane such that $\psi=0$ (though, as discussed in the appendix, the Wigner angle is
identically zero in this case when both the photon and observer geodesics are in the equatorial plane).
Finally, note that nowhere in the above argument have we made use of
the fact that $\m{k}$ is a null vector.

\section{Entanglement considerations}
\subsection{Photon Helicity States}
As discussed in Section~\ref{CST_theory}, in CST the transformation of a photon state $\ket{\m{k},\lambda}$
of pure helicity $\lambda$ has the same form as the SR flat spacetime form
\be{entneg1}
U(\Lambda)\,\ket{\m{k},\lambda} = e^{i\,\lambda\,\psi(\Lambda,\vec{n})}\ket{\m{k}',\lambda},
\ee
if we interpret $\m{k}=(k^{\ho},|\vec{k}|\vec{n})\,$,  ($\vec{n}\cdot\vec{n}=1$) as the the photon wavevector as measured by the
observer described by the tetrad $\bfe{a}(x)$, i.e. with components
$\m{k}\leftrightarrow k^{\ha}(x) = \inveT{a}{\a}(x)\, k^{\a}(x)$, and $\Lambda$ a LLT as given in \Eq{12}
\Bea
k^{\ha}(x) &\to& k'^{\ha}(x) \equiv k^{\ha}(x) + \delta k^{\ha}(x)  \no
&=& \Lambda^{\ha}_{\spp \hb}(x)\, k^{\hb}(x) =
\left( \delta^{\ha}_{\spp \hb} + \lambda^{\ha}_{\spp \hb}(x)\,d\xi \right)\, k^{\hb}(x),
\Eea
for the infinitesimal motion of the photon along its geodesic from $x^\u \to x^{'\u} = x^\u + k^\u(x)\,\d\xi$.
As discussed earlier, the Wigner rotation angle is a function of the LLT  $\lambda^{\ha}_{\spp \hb}(x)$, the
direction of propagation $\vec{n}$ of the photon, but not its frequency $\omega = |\vec{k}|$,
i.e. $\psi = \psi(\Lambda,\vec{n})$.

\Eq{entneg1} holds for an infinitesimal LLT with $\psi(\Lambda,\vec{n})$ computed from \Eq{21} and \Eq{25}.
For finite motion of the photon along its geodesic from $x^\u \to x^{'\u}$ the time ordered
expression for the the Wigner angle in \Eq{26} must be employed, and \Eq{entneg1} generalized to
\bea{ent0}
\lefteqn{U(\Lambda)\,\ket{k^{\ha}(x),\,\lambda} =
T\, [e^{i\lambda\int \tilde{\psi}(\Lambda(\xi),\,\vec{n}(\xi))d\xi}] }\no
&\times& \ket{\,T\,[e^{\int \tilde{\lambda}^{\ha}_{\spp \hb}(\Lambda(x(\xi)),\,k(\xi))d\xi}]\,k^{\hb}(x),\,\lambda}.
\eea
The term outside the ket is the time ordered product of the infinitesimal Winger rotation
$\tilde{\psi}(\Lambda(\xi),\,\vec{n}(\xi))$ along the photon's geodesic parameterized by $\xi$, while the term
inside the ket is the time ordered product of the LLT
$\tilde{\lambda}^{\ha}_{\spp \hb}\big(\Lambda(x(\xi)),\,k(\xi)\big)$ along the observer's geodesic
through which the photon passes. The time ordered integration is over the extent of $\xi$ for which
the photon passes through the observer's local laboratory.

A more relevant way to interpret \Eq{ent0} is as follows. Rather than considering a single observer with
tetrad $\bfe{a}(x)$ and following his motion through spacetime , we consider $\bfe{a}(x)$ as
describing an infinite set of observers distributed throughout spacetime at points $x$ (e.g. the
class of stationary observers with tetrad \Eq{29}, each at a fixed spatial location). Then, as the
photon passes through the local laboratories of this set of observers with tetrad $\bfe{a}(x)$,
$\tilde{\lambda}^{\ha}_{\spp \hb}\big(\Lambda(x(\xi)),\,k(\xi)\big)$ describes the
LLT applied to each local laboratory at $x$, as the observer measures the local 4-momentum $k^{\ha}(x)$
of the photon at $x$. Thus, the evolution of the photon helicity state in \Eq{ent0} is that measured
by this class of observers (vs a single observer).

Suppose we have a bipartite photon helicity Bell state at the spacetime point $x$ of the form
\be{ent1}
\ket{\Phi(x)} = \ket{k^{\ha}_1(x),\lambda_1} \, \ket{k^{\hb}_2(x),\lambda_2} \pm
\ket{k^{\ha}_1(x),\lambda_2} \, \ket{k^{\hb}_2(x),\lambda_1}.
\ee
As a specific example, one might consider two photons of helicity $\lambda_1$ and $\lambda_2$,
one traveling inward and the other traveling outward along a common geodesic (e.g. a $\pm\,k^r(x)$ in \Eq{37}).

The infinitesimal evolution of this state along the photon's trajectories is then
\bea{ent2}
\lefteqn{U(\Lambda)\,\ket{\Phi(x)} = e^{i\lambda_1\,\psi(\Lambda,\vec{n}_1)}\,\ket{k'^{\ha}_1(x),\lambda_1} \,
                              e^{i\lambda_2\,\psi(\Lambda,\vec{n}_2)}\,\ket{k'^{\hb}_2(x),\lambda_2} }\no
&\pm&  e^{i\lambda_2\,\psi(\Lambda,\vec{n}_1)}\,\ket{k'^{\ha}_1(x),\lambda_2} \,
       e^{i\lambda_1\,\psi(\Lambda,\vec{n}_2)}\,\ket{k'^{\hb}_2(x),\lambda_1}, \no
&=&  \ket{k'^{\ha}_1(x),\lambda_1} \,\ket{k'^{\hb}_2(x),\lambda_2} \no
&\pm&  e^{-i(\lambda_1-\lambda_2)\,(\psi(\Lambda,\vec{n}_1)-\psi(\Lambda,\vec{n}_2))}\,
      \ket{k'^{\ha}_1(x),\lambda_2} \, \ket{k'^{\hb}_2(x),\lambda_1},\no
\eea
where in the last line we have dropped an overall phase.
As we let these photons separate macroscopically we have to apply the time ordering operations as in \Eq{ent0}
\begin{widetext}
\bea{ent3}
U(\Lambda)\,\ket{\Psi(x)} &=&
T\, [e^{i\lambda_1\int \tilde{\psi}(\Lambda(\xi),\,\vec{n}_1(\xi))d\xi}]\,
\ket{\,T\,[e^{\int \tilde{\lambda}^{\ha}_{\spp \hc}(\Lambda(\xi),\,k(\xi))d\xi}]\,k^{\hc}_1(x),\lambda_1} \;
T\, [e^{i\lambda_2\int \tilde{\psi}(\Lambda(\xi),\,\vec{n}_2(\xi))d\xi}]\,
\ket{\,T\,[e^{\int \tilde{\lambda}^{\hb}_{\spp \hd}(\Lambda(\xi),\,k(\xi))d\xi}]\,k_2^{\hd}(x),\lambda_2}\no
&\pm&
T\, [e^{i\lambda_2\int \tilde{\psi}(\Lambda(\xi),\,\vec{n}_1(\xi))d\xi}]\,
\ket{\,T\,[e^{\int \tilde{\lambda}^{\ha}_{\spp \hc}(\Lambda(\xi),\,k(\xi))d\xi}]\,k^{\hc}_1(x),\lambda_2} \;
T\, [e^{i\lambda_1\int \tilde{\psi}(\Lambda(\xi),\,\vec{n}_2(\xi))d\xi}]\,
\ket{\,T\,[e^{\int \tilde{\lambda}^{\hb}_{\spp \hd}(\Lambda(\xi),\,k(\xi))d\xi}]\,k_2^{\hd}(x),\lambda_1}\no
\eea
\end{widetext}
The lesson of the above expression is that over some macroscopic (non-infinitesimal) evolution of the photon trajectories,
the relative phase between the product photon states (as in \Eq{ent2}) depends on the class of local laboratories
(tetrads) of the observers that the photons pass through. That is, we need to know the set of observers along
the trajectory of the photons in order to determine the relative phase, since the state of motion of
these observers (tetrads) determines the locally measured components of the direction of the photon $n_1(x(\xi))$
and $n_2(x(\xi))$, and hence the observer measured infinitesimal Wigner rotation angles
$\tilde{\psi}(\Lambda(\xi),\,\vec{n}_j(\xi))$ at each spacetime point $x(\xi)$.
For example, as discussed in Section~\ref{examples_CST}, for any motion of both the
photon and observer in the equatorial plane,
the phase factor in \Eq{ent2} is unity and the Bell state in \Eq{ent2} and \Eq{ent3}
retains its original form of \Eq{ent1}.

\subsection{More General Photon States}
In the above we have considered pure photon helicity states with a definite 4-momentum $\m{k}\leftrightarrow k^{\ha}$.
In general, we can form wave packets states over a distribution of 4-momenta, and a linear combination of
the helicities. The most general photon state has the form \cite{tung}
\bea{ent4}
\lefteqn{\ket{\psi} = \sum_{\lambda = \pm 1} \int \tilde{d}k \, \psi^\lambda(k) \, \ket{\m{k},\lambda},}\\
\tilde{d}k &=& \frac{1}{(2\pi)3}\,\theta(k^{\hat{0}}) \delta^{(3)}(\vec{k}-\vec{k}')\,\delta^{\lambda'}_\lambda
= \frac{d^3k}{(2\pi)^3 2 k^{\hat{0}}}, \no
& & \sum_{\lambda = \pm 1} \int \tilde{d} k \, |\psi^\lambda(k)|^2  = 1. \no
\langle \m{k}',\lambda'\ket{\m{k},\lambda} &=& (2\pi)^3\,2k^{\ho}\,\delta^{(3)}(\vec{k}-\vec{k}')\,\delta^{\lambda'}_{\spp \lambda}
\equiv  \tilde{\delta}^{(3)}(\vec{k}-\vec{k}')\,\delta^{\lambda'}_{\spp \lambda}, \nonumber
\eea
where $\tilde{d}k$ is the invariant Lorentz integration measure,  and we have used the
covariant normalization convention for the inner product of helicity states.
A linear polarization state (LPS) of definite 4-momentum $\m{k} = (k^{\hat{0}}, |\vec{k}|\,\vec{n})$
(where $\omega = |\vec{k}|$) and polarization angle $\phi$ is given by a linear combination
of the two photon helicity states of the form
\be{ent5}
\ket{\m{k},\phi} \equiv \ket{(k^{\hat{0}}, |\vec{k}|\,\vec{n}),\phi} =
\frac{1}{\sqrt{2}} \, \sum_{\lambda = \pm 1} \, e^{i\lambda\phi}\,\ket{\m{k},\lambda}.
\ee
In \Eq{ent5} the polarization angle $\phi$ is defined (as described in detail in the appendix) in
the \tit{standard} photon frame in which the photon propagates along the $\hat{\mathbf{z}}$-axis
and the transverse polarization vectors lie in the $\hat{\mathbf{x}}$-$\hat{\mathbf{y}}$ plane, one with
angle $\phi$ with respect to the $\hat{\mathbf{x}}$-axis, and the other with angle $\phi+\pi/2$.
The polarization angle $\phi$ is independent of the photon 4-momentum.

The most general LPS with definite direction $\vec{n}$ is given by the photon wave packet
\be{ent6}
\ket{g,\phi,\vec{n}} = \frac{1}{\sqrt{2}} \, \sum_{\lambda = \pm 1} \, e^{i\lambda\phi}\,
\int d|\vec{k}| \, g(|\vec{k}|)\, \ket{(k^{\hat{0}}, |\vec{k}|\,\vec{n}),\lambda}.
\ee
We can form a reduced helicity density matrix of the state $\ket{g,\phi,\vec{n}}$ by
projecting out the momenta
\bea{ent7}
\lefteqn{\rho_{red}(g,\phi,\vec{n}) = \int \tilde{d} k \langle\m{k}\ket{g,\phi,\vec{n}}\bra{g,\phi,\vec{n}} \m{k} \rangle} \no
&=& \sum_{\lambda\lambda'} \half e^{i(\lambda-\lambda')\phi}\ket{\lambda}\bra{\lambda'} =
\half\left(
  \begin{array}{cc}
    1 & e^{i 2\phi} \\
    e^{-i 2\phi} & 1 \\
  \end{array}
\right),
\eea
where  the rows and columns of the matrix are labeled by the helicity index in the order $\lambda = \{1,-1\}$.

In flat spacetime, Caban and Rembielinski \cite{rembielinski} showed that under a SR LT this LPS remains
a LPS, with a new polarization angle $\phi' = \phi + \psi(\Lambda,\vec{n})$ via
\bea{ent8}
\lefteqn{U(\Lambda)\ket{g,\phi,\vec{n}} = \ket{g',\phi,\vec{n}'},\quad g'(|\vec{k}|)
= \frac{2}{a} g\left( \frac{2 |\vec{k}|}{a} \right),} \no
&=& \frac{1}{\sqrt{2}} \, \sum_{\lambda = \pm 1} \, e^{i\lambda(\phi+\psi(\Lambda,\vec{n}))}\,
\int d|\vec{k}| \,g'(|\vec{k}|)\, \ket{(k^{\hat{0}}, |\vec{k}|\,\vec{n}'),\lambda}, \no
\eea
where $\vec{n}$ and $g(|\vec{k}|)$  are the original photon propagation direction and
wave packet momentum distribution, and $\vec{n}'$ and $g'(|\vec{k}|)$ are the
corresponding transformed quantities. The authors showed that the reduced
helicity density matrix for transformed state $U(\Lambda)\ket{g,\phi,\vec{n}} $
transforms properly under LTs, i.e.
\bea{ent9}
\lefteqn{\rho'_{red} = U(\Lambda)\rho_{red}(g,\phi,\vec{n})U^\dagger(\Lambda)} \no
&=&\half\left(
  \begin{array}{cc}
    1 & e^{i 2(\phi+\psi(\Lambda,\vec{n}))} \\
    e^{-i 2(\phi+\psi(\Lambda,\vec{n}))} & 1 \\
  \end{array}
\right)\no
&=& \rho_{red}(g',\phi+\psi(\Lambda,\vec{n})),\vec{n}'), \quad U(\Lambda)_{\lambda \lambda'} = e^{i\lambda\psi(\Lambda,\vec{n})}\delta_{\lambda\lambda'}\no
\eea
The fact that the LPS admits a covariant description of the reduced density matrix
in terms of helicity degrees of freedom is related to the fact that LTs do not create
entanglement between the helicity and momentum directions \cite{rembielinski}, as indicated by the matrix
elements of $U(\Lambda)$ in the last line of \Eq{ent9} arising from fundamental transformation
law \Eq{entneg1}. This is very different from the situation for massive particles
in which the action of $U(\Lambda)$ \Eq{1} does entangle momentum and spin \cite{adami_spin}.
These considerations remain true in CST, where by the EP,
SR holds locally at each spacetime point if we interpret $k^{\ha}(x) = \inveT{a}{\a}(x)\,k^\a(x)$
as the components of the photon 4-momentum as measured by an observer at $x$ with tetrad $\bfe{a}(x)$.

The transformation of $g\to g'$ in \Eq{ent8} arose from
applying the unitary transformation \Eq{entneg1} to the helicity states, making a
change of integration variable from $|\vec{k}|$ to $|\vec{k}'|$ and using the frequency
transformation law $k'^{\ho} = a k^{\ho}/2$ from \Eq{20.5} (with a final relabeling
of $|\vec{k}'|$ to $|\vec{k}|$).
Note that the transformed polarization angle $\phi' = \phi + \psi(\Lambda,\vec{n})$
involves the Wigner angle $\psi(\Lambda,\vec{n})$ evaluated at the original propagation
direction of the photon $\vec{n}$, while the kets on the right hand side are evaluated at
the transformed photon direction $\vec{n}'$. Since $\psi(\Lambda,\vec{n})$ depends only
on the direction of the photon and not on its frequency $\omega = |\vec{k}|$, the
phase factor $e^{i\lambda\,\psi}$, resulting from the unitary transformation of the photon helicity states,
can be pulled outside the integral.

A general two particle photon state takes the form
\bea{ent10}
\ket{\Psi} &=& \sum_{\lambda_1,\lambda_2}\int \int \tilde{d} k_1 \tilde{d} k_2 \, g_{\lambda_1\lambda_2}(\m{k}_1,\m{k}_2)
  \ket{\m{k}_1,\lambda_1} \, \ket{\m{k}_2,\lambda_2}, \no
& & \sum_{\lambda_1,\lambda_2}\int \int \tilde{d} k_1 \tilde{d} k_2 \, |g_{\lambda_1\lambda_2}(\m{k}_1,\m{k}_2)|^2=1.
\eea
For the choice of the distribution function $g_{\lambda_1\lambda_2}(\m{k}_1,\m{k}_2)$
\be{ent11}
g_{\lambda_1\lambda_2}(\m{k}_1,\m{k}_2) = \frac{1}{\sqrt{2}} \, e^{i\lambda_1\phi_1} \, e^{i\lambda_2\phi_2} \,
\delta_{\lambda_1\lambda_2} f(\m{k_1},\m{k_2}),
\ee
where $\phi_1$ and $\phi_2$ constant, momentum independent polarization angles, the state
$\ket{\Psi}$ is fully entangled in helicity and entangled in momentum if $f$ is non-factorizable, i.e.
$f(\m{k_1},\m{k_2})\ne f_1(\m{k}_1)\,f_2(\m{k}_2)$. In analogy with the single particle state \Eq{ent7},
the helicity reduced density matrix obtained by tracing the pair of momenta is
\be{ent12}
\rho_{red}(\phi_1,\phi_2)= \half\,
\left(
  \begin{array}{cccc}
    1 & 0 & 0 & e^{i2(\phi_1+\phi_2)} \\
    0 & 0 & 0 & 0 \\
    0 & 0 & 0 & 0 \\
    e^{-i2(\phi_1+\phi_2)}& 0 & 0 & 1 \\
  \end{array}
\right),
\ee
where the normalization condition for $g_{\lambda_1\lambda_2}$  in \Eq{ent10} has been used,
and the rows and columns of the matrix are labeled by the double helicity indices
$\lambda_1\lambda_2 = \{11,\,1-1,\,-1 1,\,-1-1\}$.
Under the action of an infinitesimal LLT $\mathcal{U}(\Lambda)=U(\Lambda)\otimes U(\Lambda)$
the reduced helicity density matrix for the state $\mathcal{U}(\Lambda)\,\ket{\Psi}$ is
\bea{ent13}
\lefteqn{\rho'_{red}(\phi_1,\phi_2)= \half\,
\left(
  \begin{array}{cccc}
    1 & 0 & 0 & (\rho'_{red})_{11,-1-1} \\
    0 & 0 & 0 & 0 \\
    0 & 0 & 0 & 0 \\
    (\rho'_{red})^*_{11,-1-1}& 0 & 0 & 1 \\
  \end{array}
\right)}, \no
& & (\rho'_{red})_{11,-1-1} = \int\int \tilde{d} k_1 \tilde{d} k_2 |f'(\m{k}_1,\m{k}_2)|^2
e^{i 2 \Phi(\vec{n}_1,\vec{n}_2)} \no
& & \Phi(\vec{n}_1,\vec{n}_2) = \phi_1+\psi(\Lambda,\vec{n}_1)+\phi_2+\psi(\Lambda,\vec{n}_2).
\eea
In \Eq{ent13} $f'$ is the transformed distribution function (analogous to $g'$ in \Eq{ent8})
that depends on the untransformed frequencies $|\vec{k}_j|$, but the transformed directions $\vec{n}'_j$.
Since $\tilde{d}k_j \propto  d^3k_j = \d\Omega_{\vec{n}_j}\,d|\vec{k}_j|\,|\vec{k}_j|^2$ involves
an integration over the untransformed photon directions $\vec{n}_j$, (where $\vec{n}'_j$ and
$\vec{n}_j$ are related by \Eq{20.5}), the integral over $\d\Omega_{\vec{n}_j}$ is in general
very complicated, mixing up the photon directions, but again without entangling with the helicity.
Without the factors of $\psi_{\vec{n}_j}$ in the argument of the phase of $(\rho'_{red})^*_{11,-1-1}$
\Eq{ent13} reduces to \Eq{ent12}, showing that analogous to \Eq{ent9},
the reduced helicity density matrix $\rho'_{red}(\phi_1,\phi_2)$
transforms covariantly under LLTs. In CST, it is possible that the observer (tetrad) changes for each
LLT along the trajectory of the photons, from which the
local Wigner angles $ \psi_{\vec{n}_i}(x(\xi))$ are measured by (massive) observers.
\section{Summary and Conclusions}
The Wigner rotation for a photon can be envisioned as the rotation of the transverse linear polarization vectors,
in the plane perpendicular to the direction of propagation of the photon, resulting from
a Lorentz transformation $\Lambda$ between observers. The natural quantum state
description of the photon is in terms of helicity states $\ket{\m{k},\lambda}$, $\lambda = \pm1$,
$\m{k} = (k^{\ho},|\vec{k}|\vec{n})$,
in which the corresponding induced unitary transformation $U(\Lambda)$ introduces a phase factor,
dependent upon $\Lambda$ and the propagation direction of the photon $\vec{n}$, without changing
the helicity, i.e. $U(\Lambda)\ket{\m{k},\lambda} = e^{i\lambda\psi(\Lambda,\vec{n})}\ket{\m{k}',\lambda}$.
In the flat spacetime of special relativity
$\Lambda$ transforms between a special class of observers, namely inertial observers for which the
acceleration of the observer is zero (constant velocity observers). Such observers are global
in the sense that they are position independent and exist over the whole of the flat spacetime.

In going to curved spacetime (CST) where general relativity applies, all types of observers,
in arbitrary states of motion, are allowed. The motion of these observers is now reduced
to a local description, encapsulated in an orthonormal tetrad $\bfe{a}(x)$ that describes the four axes (three
spatial and one temporal) that defines the observer's local laboratory at the spacetime point $x$
from which he makes measurements. For example, the photon 4-momentum $k^\a(x)$ existing in
a CST described by coordinates $x^\a$, has components $k^{\ha}(x)$ in the observer's local
laboratory given by $k^{\ha}(x) = \inveT{a}{\a}(x)\,k^{\a}(x)$.

By the equivalence principle, the laws of special relativity
apply in this local laboratory (local tangent plane to the curved spacetime), at the spacetime point $x$.
Therefore, we can compute the Wigner rotation angle in CST by applying the calculational procedure
appropriate for flat spacetime to the observer's instantaneous local laboratory. The quantum
state of the photon is described by the local helicity state $\ket{k^{\ha}(x),\lambda}$ and
the observer by the tetrad $\bfe{a}(x)$. The instantaneous Wigner rotation angle $\psi(\Lambda,\vec{n})$,
arising from a local Lorentz transformation as the photon traverses infinitesimally along its geodesic,
now depends  on the propagation direction of the photon $(\vec{n})^{\hi} = k^{\hi}/|\vec{k}|$
as measured locally by the observer at $x$.

In this work we have developed the local Wigner rotation for photons in an arbitrary CST. We have given
specific examples in the case of Schwarzschild spacetime and compared these with the results
from flat spacetime. The difference in the CST case is that an explicit description of the
observer, via his tetrad, is needed to compute the local Wigner angle. That is, the locally measured
Wigner rotation angle is observer dependent, which we develop explicitly. In terms of a local helicity state
description of the quantum photon states, the induced local Lorentz transformation that
gives rise to the local Wigner angle as the photon traverses its geodesic, does not entangle
the photon direction with the helicity, in contrast to the case for spin-momentum entanglement
that occurs for massive particles. We have also developed a sufficient condition for observers
who would measure zero Wigner rotation and have shown that such observers can be in Fermi-Walker
frames, i.e. the instantaneous non-rotating rest frame of the accelerating observer.

\begin{acknowledgments}
PMA wishes to acknowledge the support of  the Air Force Office of Scientific Research (AFOSR) for this work.
\end{acknowledgments}

\appendix
\label{app}

\section{Wigner rotation in flat spacetime: examples}
In this appendix we give explicit examples illustrating the operational meaning of the Wigner rotation in
flat spacetime in terms of its effect on the polarization vectors for photons.

As given in \Eq{6}, the polarization vectors for positive and negative helicity states
$\epsilon^\u_\pm(\hat{\mbf{k}})$ (right and left circular polarization) with
propagation 4-vector $\m{k}=(k^{\ho},|\vec{k}|\,\vec{n})$ are given by
\be{A1}
\epsilon^\u_\pm(\hat{\mbf{k}}) = \frac{R(\hat{\mbf{k}})}{\sqrt{2}} \,
\left[
\begin{array}{c}
  0 \\
  1 \\
  \mp i \\
  0
\end{array}
\right],
\ee
with the components of the column vector labeled by the Cartesian coordinates $x^\u=(t,x,y,z)$.
Here $R(\hat{\mbf{k}})$ is the rotation that takes the \tit{standard} direction $\hat{\mathbf{z}}$-axis to the
photon propagation direction $\hat{\mbf{k}}= \vec{k}/|\vec{k}|\,(=\vec{n})$.
Under a LT $\Lambda$, the polarization vector transforms as $\epsilon_\pm^{\u}\to\epsilon^{'\u}$
with \cite{adami_photon}
\bea{A2}
\epsilon_\pm^{'\u}(\hat{\mbf{k}'}) &\equiv& D(\Lambda)\,\epsilon_\pm^\u(\hat{\mbf{k}}) \no
&=& R(\Lambda \hat{\mbf{k}}) \, R_z(\psi(\Lambda,\vec{n})) \,
                          R(\hat{\mbf{k}})^{-1} \, \epsilon_\pm^\u(\hat{\mbf{k}}), \quad \\
&=& \Lambda  \epsilon_\pm^\mu(\hat{\mbf{k}}) - \frac{(\Lambda\,\epsilon_\pm^\mu(\hat{\mbf{k}}))^0}{(\Lambda\,k^\mu)^0} \, \Lambda  k^\mu.
\label{A3}
\eea
Here we use the typical ``abuse of notation" denoting $\Lambda\,\hat{\mbf{k}}$ for the transformed photon direction
$\hat{\mbf{k}}' = \vec{k}'/|\vec{k}'|$, where  $\vec{k}'$ is the 3-vector portion of
the transformed photon 4-momentum $k^{'\mu} = \Lambda^{\mu}_{\spp \nu}\, k^\nu$. Thus
$R(\Lambda\,\hat{\mbf{k}})$ is the rotation taking the standard direction $\hat{\mathbf{z}}$ to $\hat{\mbf{k}}'$.

From \Eq{A1} we can construct a linear polarization vector (LPV) $\epsilon^\mu_\phi(\hat{\mbf{k}})$
\bea{A4}
\epsilon^\mu_\phi(\hat{\mbf{k}}) &=& \frac{1}{\sqrt{2}}\,
   \left( e^{i \phi}\, \epsilon^\mu_+(\hat{\mbf{k}}) +  e^{-i \phi}\, \epsilon^\mu_-(\hat{\mbf{k}}) \right), \no
&=& R(\hat{\mbf{k}}) \,
\left[
\begin{array}{c}
  0 \\
  \cos\phi \\
  \sin\phi \\
  0
\end{array}
\right] \equiv R(\hat{\mbf{k}}) \, \tilde{\epsilon}^\mu_\phi(\hat{\mathbf{z}}),
\eea
where the polarization angle $\phi$ is defined by the angle the LPV $\tilde{\epsilon}^\mu_\phi(\hat{\mathbf{z}})$
makes with the $\hat{\mathbf{x}}$-axis when the photon propagates along the $\hat{\mathbf{z}}$-axis
in this \tit{standard frame}, i.e. $\hat{\tilde{\mathbf{k}}}=\hat{\mathbf{z}}$ (denoted by \tit{tildes} over vectors).
The LPV in the standard frame $\tilde{\epsilon}^\mu_\phi(\hat{\mathbf{z}})$ is obtained by rotating the polarization vector
$\epsilon^\mu_\phi(\hat{\mbf{k}})$ propagating in the direction $\hat{\mbf{k}}$, by the transformation
that takes  $\hat{\mbf{k}}$ back to $\hat{\mathbf{z}}$, i.e. by the rotation $R^{-1}(\hat{\mbf{k}})$.

After a LT $\Lambda$, the new LPV is given by
\bea{A5}
\epsilon'^{\mu}_{\phi'}(\hat{\mbf{k}}') &=& \frac{1}{\sqrt{2}}\,
   \left( e^{i \phi'}\, \epsilon'^{\mu}_+(\hat{\mbf{k}'}) +  e^{-i \phi'}\, \epsilon'^{\mu}_-(\hat{\mbf{k}'}) \right), \no
&=& R(\Lambda\,\hat{\mbf{k}}) \,
\left[
\begin{array}{c}
  0 \\
  \cos\phi' \\
  \sin\phi' \\
  0
\end{array}
\right] \equiv R(\Lambda\,\hat{\mbf{k}}) \, \tilde{\epsilon}'^{\mu}_{\phi'}(\hat{\mathbf{z}}).
\eea
Again, the transformed LPV $\tilde{\epsilon}'^\mu_{\phi'}(\hat{\mathbf{z}})$ in the standard frame
($\hat{\tilde{\mathbf{k}}}'=\hat{\mathbf{z}}$) is obtained by rotating the polarization vector
$\epsilon'^\mu_{\phi'}(\Lambda\,\hat{\mbf{k}})$ propagating in the direction $\hat{\mbf{k}}'$, by the rotation
that takes  $\hat{\mbf{k}}'$ back to $\hat{\mathbf{z}}$, i.e.  $R^{-1}(\Lambda\,\hat{\mbf{k}})$.
The angle  that $\tilde{\epsilon}'^\mu_{\phi'}(\hat{\mathbf{z}})$ makes with the $\hat{\mathbf{x}}$-axis
in the standard frame defines the transformed polarization angle $\phi'$.

By multiplying \Eq{A2} by $R^{-1}(\Lambda\,\hat{\mbf{k}})$ and inserting \Eq{A4} and \Eq{A5} we obtain
\be{A6}
\tilde{\epsilon}'^{\mu}_{\phi'}(\hat{\mathbf{z}})=R_z(\psi(\Lambda,\vec{n})) \; \tilde{\epsilon}^\mu_\phi(\hat{\mathbf{z}}),
\ee
which upon comparing the arguments of the trigonometric functions yields
\be{A7}
\phi' = \phi + \psi(\Lambda,\vec{n}).
\ee
This states that the effect of a LT $\Lambda$ is a rotation of the polarization angle $\phi\to\phi'$
in the standard frame  by the Wigner angle $\psi(\Lambda,\vec{n})$, i.e. a rotation of the \tit{standard} polarization
vectors once we bring them back to the \tit{standard frame}  in which the photon propagates along the
$\hat{\mathbf{z}}$-axis, ($\hat{\tilde{\mathbf{k}}}=\hat{\mathbf{z}}$), and the standard polarization vectors
lie in the $\hat{\mathbf{x}}$-$\hat{\mathbf{y}}$ plane.
It is in this standard frame that we most easily measure the polarization angles $\phi$ and $\phi'$ and therefore determine
the Wigner angle $\psi(\Lambda,\vec{n})$. In the following we illustrate a few sample cases in which the Wigner angle is zero,
and non-zero.

\begin{figure}[h]
\includegraphics[width=3.75in,height=2.75in]{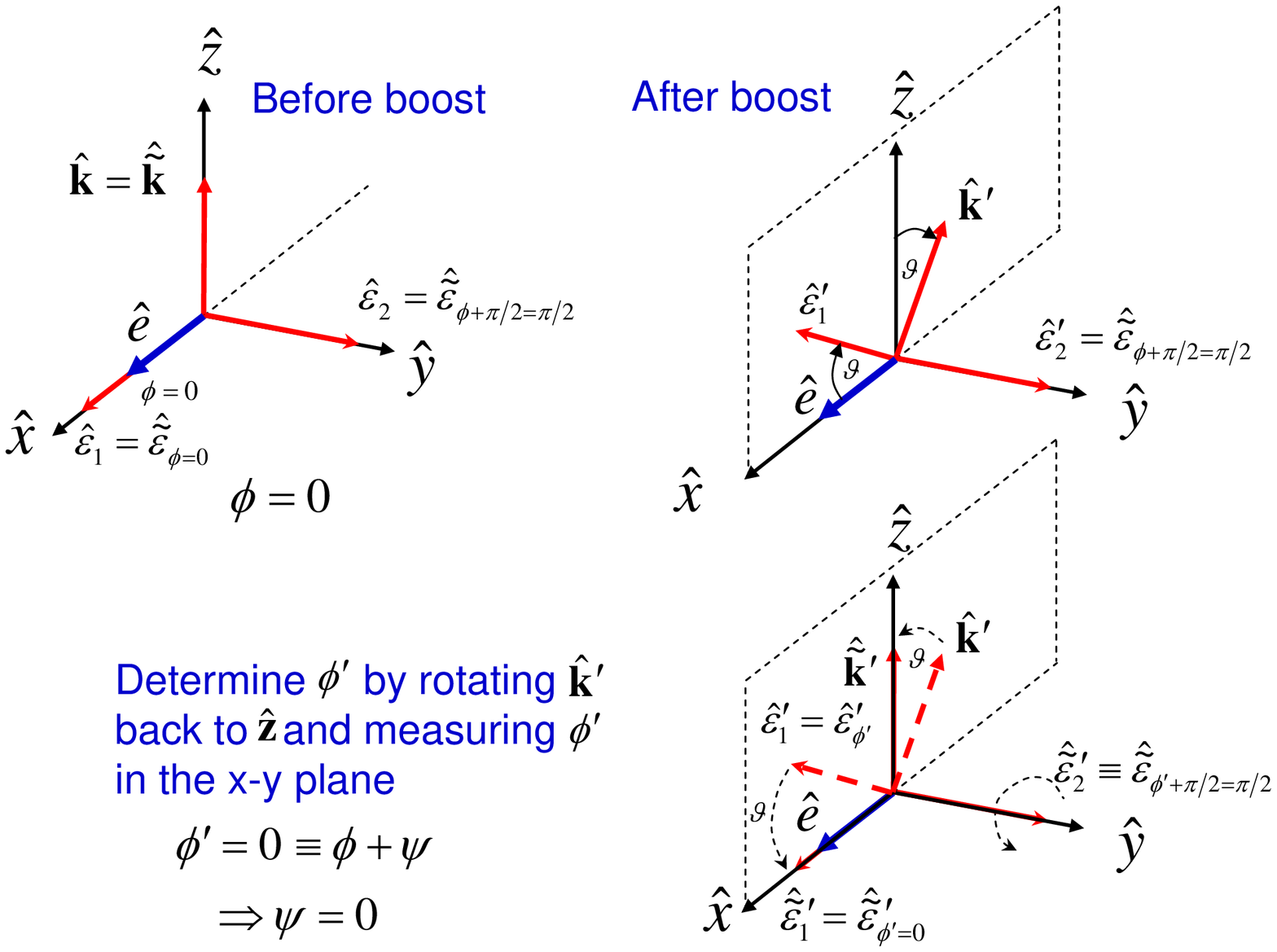}
\caption{(Color online) Example of a zero Wigner angle $\psi$: photon propagation direction $\hat{\mbf{k}}$ along the $\mathbf{z}$-axis,
boost $\hat{\mathbf{e}}$ along the $\mathbf{z}$-axis.
In fact, the Wigner angle $\psi$ is identically zero for any boost direction $\hat{\mathbf{e}}$
if the photon $\hat{\mbf{k}}$ travels along the $\mathbf{z}$-axis.}
\label{figA1}
\end{figure}

In \Fig{figA1} we consider the photon to be traveling along the $\hat{\mathbf{z}}$-axis, and consider a boost
along the $\hat{\mathbf{x}}$-axis. We denote the boost direction by $\hat{\mathbf{e}}\equiv \vec{v}/|\vec{v}|$
where $\vec{v}$ is the velocity of the frame we are transforming to and $\xi$ defined by $\tanh\xi = -|\vec{v}|/c$
is the rapidity parameter of the boost.
We take the two transverse polarization vectors to lie along the
$\hat{\mathbf{x}}$ and $\hat{\mathbf{y}}$ axes, so that the $\hat{\mathbf{x}}$ polarization vector
has a polarization angle $\phi=0$. The boost rotates the propagation vector $\hat{\mbf{k}}$ and the $\hat{\mathbf{x}}$
polarization vector in the $\hat{\mathbf{x}}$-$\hat{\mathbf{z}}$ plane counterclockwise about the  $\hat{\mathbf{y}}$-axis
by some boost dependent angle $\vartheta$ \cite{alsing_milburn,adami_photon},
leaving the $\hat{\mathbf{y}}$ polarization vector  unchanged. To determine $\phi'$ we rotate $\hat{\mbf{k}}'$ back
to the $\hat{\mathbf{z}}$-axis clockwise about the $\hat{\mathbf{y}}$-axis, undoing the original rotation, and thus returning
all the vectors to their original orientations. Therefore, $\phi'= \phi$ and thus the Wigner angle is zero.
It is straightforward to show that for the photon traveling along the $\hat{\mathbf{z}}$-axis, a boost along \tit{any}
direction yields a zero Wigner angle.

In general for $\hat{\mbf{k}}$ not along $\hat{\mathbf{z}}$,
we can determine the polarization angles $\phi$ and $\phi'$ as follows. Given a photon
propagating along the direction $\hat{\mbf{k}}$ with polarization vectors in a plane perpendicular to this vector,
we find the polarizations vectors in the standard frame by applying the rotation
$R_{\hat{\mbf{k}}\times\hat{\mathbf{z}}}(\theta)$ to the triad, where $\cos\theta = \hat{k}_3$ which takes
$\hat{\mbf{k}}\to\hat{\mathbf{z}}$.
A pure boost along the $\hat{\mathbf{e}}$ rotates the photon propagation
direction by a boost dependent angle $\vartheta$ counterclockwise about the axis $\hat{\mathbf{e}}\times\hat{\mbf{k}}$,
$\;\hat{\mbf{k}}'=R_{\hat{\mathbf{e}}\times\hat{\mbf{k}}}(\vartheta)\,\hat{\mbf{k}}'$. To determine
the transformed polarization angle $\phi'$ we rotate $\hat{\mbf{k}}'$ counter clockwise along
the direction $\hat{\mbf{k}}'\times\hat{\mathbf{z}}$ by the angle $\theta'$ where $\cos\theta' = \hat{k}'_3$,
which takes $\hat{\mbf{k}}'\to\hat{\mathbf{z}}$, and the transformed polarization vectors to the
$\hat{\mathbf{x}}$-$\hat{\mathbf{y}}$ plane.

From the above discussion we can also infer that the Wigner angle is zero if  $\hat{\mathbf{e}}$ and
$\hat{\mbf{k}}$ both lie in the $\hat{\mathbf{x}}$-$\hat{\mathbf{z}}$ plane, or both lie in the
$\hat{\mathbf{y}}$-$\hat{\mathbf{z}}$ plane. In the latter case, if the first polarization vector also lies in
the $\hat{\mathbf{y}}$-$\hat{\mathbf{z}}$ plane orthogonal to $\hat{\mbf{k}}$, and the second polarization
vector lies along the $\hat{\mathbf{x}}$-axis, the a boost in the $\hat{\mathbf{y}}$-$\hat{\mathbf{z}}$ plane will
rotate the triad $(\hat{\mbf{k}}, \hat{\mbf{\epsilon}}_1, \hat{\mbf{\epsilon}}_2$=$\hat{\mathbf{x}})$ about
the $\hat{\mathbf{x}}$-axis. Therefore, when we undo this rotation about the $\hat{\mathbf{x}}$-axis in order to
calculate $\phi'$, the triad is returned to its original orientation, so that $\phi'=\phi$, implying a zero Wigner angle,
analogous to \Fig{figA1}.

\begin{figure}[h]
\includegraphics[width=3.5in,height=1.5in]{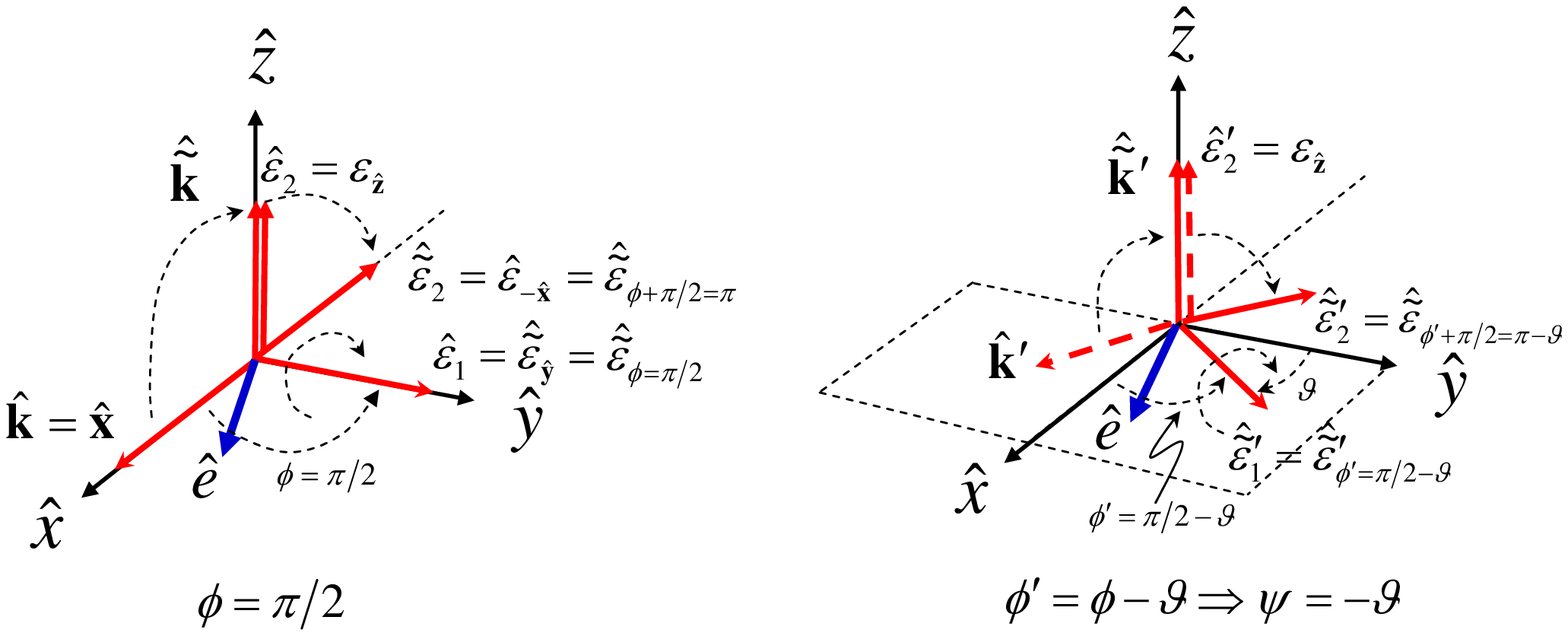}
~\caption{(Color online) Example of a non-zero Wigner angle $\psi$: photon propagation direction
$\hat{\mbf{k}}$ along the $\mathbf{x}$-axis,
boost $\hat{\mathbf{e}}$ in the $\hat{\mathbf{x}}$-$\hat{\mathbf{y}}$ plane.}
\label{figA2}
\end{figure}

However, if $\hat{\mathbf{e}}$ and $\hat{\mbf{k}}$ both lie in the $\hat{\mathbf{x}}$-$\hat{\mathbf{y}}$ plane,
the Wigner angle is non-zero, as illustrated in \Fig{figA2} for the case of
$(\hat{\mbf{k}}$=$\hat{\mathbf{x}}, \hat{\mbf{\epsilon}}_1$=$\hat{\mathbf{y}}, \hat{\mbf{\epsilon}}_2$=$\hat{\mathbf{z}})$.
Here the initial polarization angle for $\hat{\mbf{\epsilon}}_1$ is $\phi=\pi/2$. For a boost in the
$\hat{\mathbf{x}}$-$\hat{\mathbf{y}}$ plane, as indicated in the figure, $\hat{\mbf{k}}$ is rotated counter clockwise
about the $\hat{\mathbf{z}}$-axis by the boost dependent angle $\vartheta$, yielding a transformed polarization
vector $\hat{\mbf{\epsilon}}'_1$ oriented at angle $\phi' = \pi/2 - \vartheta$ with respect to the
$\hat{\mbf{x}}$-axis. Upon rotating $\hat{\mbf{k}}'$ (about $\hat{\mbf{\epsilon}}'_1$) to the
$\hat{\mathbf{z}}$-axis, $\hat{\mbf{\epsilon}}'_1$ is left invariant, maintaining the relation
$\phi' = \pi/2 - \vartheta$, or equivalent, a Wigner angle of $\psi(\Lambda,\vec{n}) = - \vartheta$.

\begin{figure}[h]
\includegraphics[width=3.75in,height=2.75in]{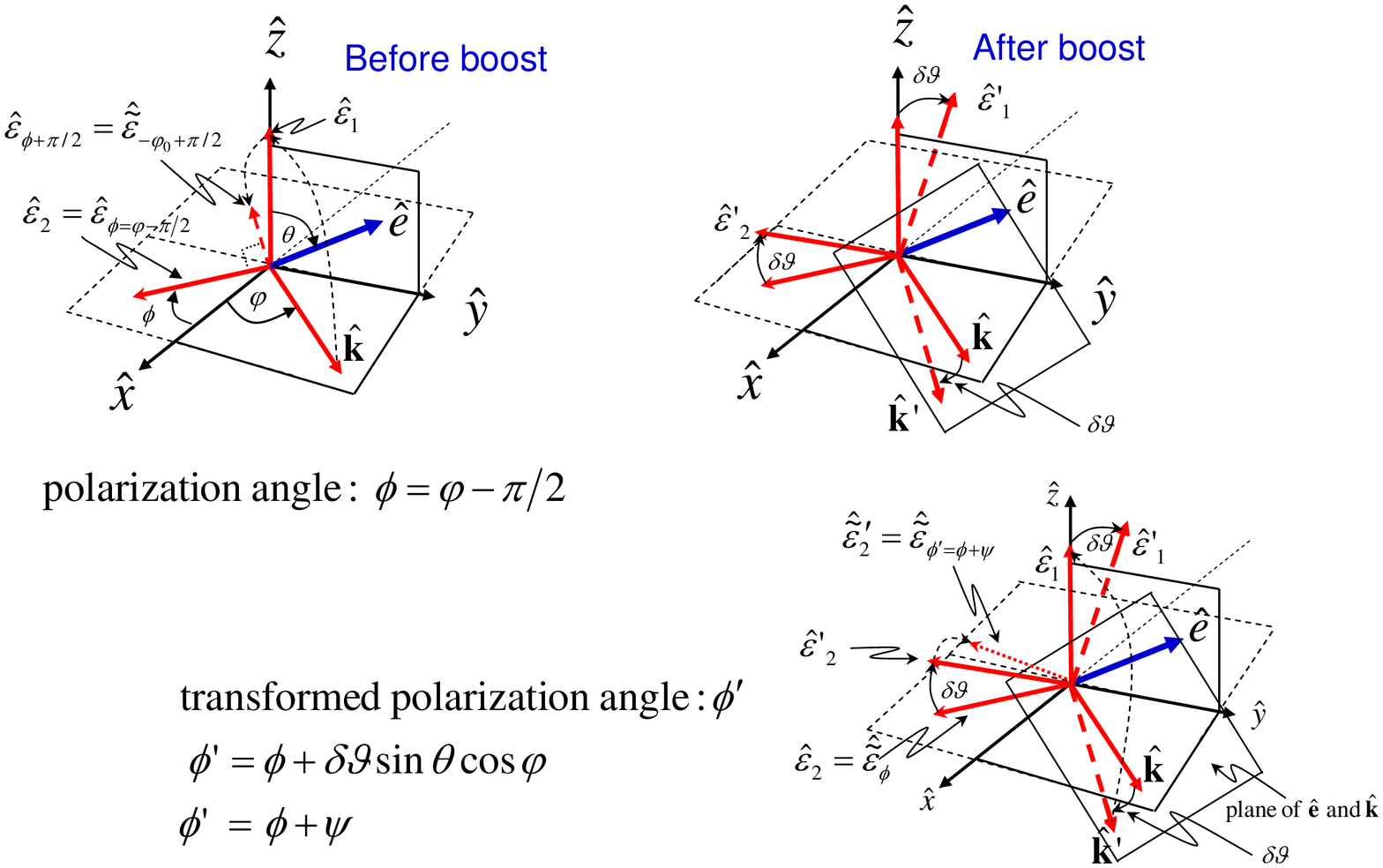}
\caption{(Color online) Second example of non-zero Wigner  angle $\psi$: photon propagation direction
$\hat{\mbf{k}}$ in the $\hat{\mathbf{x}}$-$\hat{\mathbf{y}}$ plane at azimuthal angle $\varphi$,
boost $\hat{\mathbf{e}}$ in the $\hat{\mathbf{y}}$-$\hat{\mathbf{z}}$ plane at polar angle $\theta$.}
\label{figA3}
\end{figure}

In \Fig{figA3} we illustrate one last case that is relevant to the discussion in the main body of the text when
we consider an example of a non-zero Wigner angle in curved Schwarzschild spacetime. Here in flat spacetime, we
consider the case when the photon direction $\hat{\mbf{k}}$ is in the $\hat{\mathbf{x}}$-$\hat{\mathbf{y}}$ plane at some
angle $\varphi$ to the $\hat{\mathbf{x}}$-axis, and the boost direction $\hat{\mathbf{e}}$ is in the
$\hat{\mathbf{y}}$-$\hat{\mathbf{z}}$ plane at some polar angle $\theta$ with respect to the $\hat{\mathbf{z}}$-axis.
We choose the first polarization vector to lie along
the $\hat{\mathbf{z}}$-axis, and the second polarization vector to lie
in the $\hat{\mathbf{x}}$-$\hat{\mathbf{y}}$ orthogonal to
$\hat{\mbf{k}}$ with polarization angle $\phi=\varphi-\pi/2$,
A boost along $\hat{\mathbf{e}}$ induces a rotation of the triad
$(\hat{\mbf{k}}, \hat{\mbf{\epsilon}}_1=\hat{\mathbf{z}},\hat{\mbf{\epsilon}}_2)$ about the axis
$(\hat{\mathbf{e}}\times\hat{\mbf{k}}$ orthogonal to the $\hat{\mbf{k}}$-$\hat{\mathbf{e}}$ plane
by some boost dependent angle $\vartheta$. This pushes $\hat{\mbf{k}}'$ below the
$\hat{\mathbf{x}}$-$\hat{\mathbf{y}}$ plane and pulls $\hat{\mbf{\epsilon}}'_2$ above the
$\hat{\mathbf{x}}$-$\hat{\mathbf{y}}$ plane. Upon rotating $\hat{\mbf{k}}'$ back to the $\hat{\mathbf{z}}$-axis
by a rotation about the direction $\hat{\mbf{k}}'\times\hat{\mathbf{z}}$, $\hat{\mbf{\epsilon}}'_2$
is returned to the $\hat{\mathbf{x}}$-$\hat{\mathbf{y}}$ plane as $\hat{\tilde{\mbf{\epsilon}}}'_2$. For
a boost of infinitesimal angle $\delta\varphi$ as illustrated in \Fig{figA3}, the polarization angle of
the transformed polarization vector is $\phi'=\phi+\psi(\Lambda,\vec{n})$ with the non-zero Wigner angle
given to $O(\delta\vartheta)$ as $\psi(\Lambda,\vec{n})=\delta\vartheta\sin\theta\cos\varphi=\delta\vartheta\,\hat{k}_3$.




\begin{thebibliography}{99}
\bibitem{note_1} The force free (zero acceleration) equations of motion of a (massive or massless) particle
can be obtained from the Euler-Lagrange equations $d/d\sigma(\partial L/\partial\dot{x}^\a) - \partial L/\partial x^\a=0$,
where $L = ds = (g_{\a\b}(x)\,\dot{x}^\a\,\dot{x}^\b)^{1/2}$ is the effective Lagrangian, and $\dot{x}^a = dx^\a/d\sigma$
with $\sigma$ the affine paramter along the particle's geodesic trajectory. If the metric is independent of the coordinate
$x^\a_0\equiv x_o$, the quantity $\partial L/\partial \dot{x}_o$ is a constant of the motion along the geodesic. The latter
expression can be written as $\mbf{\zeta}\cdot\m{u}=\trm{constant}$ where $u^\a = \dot{x}^\a$ is the 4-velocity
of the particle, i.e. the tangent to its geodesicc, and $\zeta^\a = \delta^{\a}_{\spp x_o}$ is
called a \tit{Killing vector}, and reprsents an isometry of the metric. The Schwarzschild metric \Eq{27}
is independent of the coordinates $t$ and $\phi$. The quantity $e=\mbf{\zeta}\cdot\m{u} = (1-r_s/r)\,\dot{t}$ for
$\zeta^\a = \delta^{\a}_{\spp t}$ represents the particle's energy (per unit mass for $m\ne 0$),
and $l=\mbf{\zeta}\cdot\m{u} = r^2\sin^2\theta\,\dot{\phi}$ for
$\zeta^\a = \delta^{\a}_{\spp \phi}$ represents the particle's orbital angular momentum
(per unit mass for $m\ne 0$).
\end{thebibliography}

\end{document}